\documentclass[12pt,preprint]{aastex}
\usepackage{lsst}
\usepackage{xspace}
\usepackage[english]{babel}
\usepackage[utf8x]{inputenc}
\usepackage{hyperref}
\usepackage{amsmath}
\usepackage{graphicx}
\usepackage{longtable}
\usepackage{comment}
\usepackage{booktabs}

\newcommand{\DIASource}{\code{DIASource}\xspace}
\newcommand{\DIASources}{\code{DIASources}\xspace}

\newcommand{\FindOrb}{\code{Find\_{}Orb}\xspace}

\begin{document}
\title{The Large Synoptic Survey Telescope as a Near-Earth Object Discovery Machine}

\author{R. Lynne Jones\altaffilmark{1},
Colin T. Slater\altaffilmark{1},
Joachim Moeyens\altaffilmark{1},
Lori Allen\altaffilmark{2},
Tim Axelrod\altaffilmark{3},
Kem Cook\altaffilmark{4},
\v{Z}eljko Ivezi\'{c}\altaffilmark{1},
Mario Juri\'{c}\altaffilmark{1},
Jonathan Myers\altaffilmark{5},
Catherine E. Petry\altaffilmark{6}
}
\altaffiltext{1}{Department of Astronomy, University of Washington, Box 351580,
  Seattle, WA 98195, USA}
\altaffiltext{2}{National Optical Astronomy Observatory, 950 North Cherry
Avenue, Tucson, AZ 85719, USA}
\altaffiltext{3}{University of Arizona, Steward Observatory, 933 North Cherry
Avenue, Tucson, AZ 85721, USA}
\altaffiltext{4}{Cook Astronomical Consulting, 220 Duxbury CT, San Ramon, CA 94583, USA}
\altaffiltext{5}{Work performed at the Large Synoptic Survey Telescope, 950 North Cherry Avenue, Tucson, AZ 85719, USA; 
Jonathan is now affiliated with Johns Hopkins University, Applied Physics Laboratory, 11100 Johns Hopkins Rd, Laurel, MD 20723, USA}
\altaffiltext{6}{Large Synoptic Survey Telescope, 950 North Cherry Avenue, Tucson, AZ 85719, USA}

\begin{abstract}
Using the most recent prototypes, design, and as-built system information, we test and quantify the capability of the Large Synoptic Survey Telescope (LSST) to discover Potentially Hazardous Asteroids (PHAs) and Near-Earth Objects (NEOs).
We empirically estimate an expected upper limit to the false detection rate in LSST image differencing, using measurements on DECam data and prototype LSST software and find it to be about $450$~deg$^{-2}$. We show that this rate is already tractable with current prototype of the LSST Moving Object Processing System (MOPS) by processing a 30-day simulation consistent with measured false detection rates.
We proceed to evaluate the performance of the LSST baseline survey strategy for  PHAs and NEOs using a high-fidelity simulated survey pointing history. 
We find that LSST alone, using its baseline survey strategy, will detect 66\% of the PHA and 61\% of the NEO population objects brighter than $H=22$, with the uncertainty in the estimate of $\pm5$ percentage points. By generating and examining variations on the baseline survey strategy, we show it is possible to further improve the discovery yields.
In particular, we find that extending the LSST survey by two additional years and doubling the MOPS search window increases the completeness for PHAs to 86\% (including those discovered by contemporaneous surveys) without jeopardizing other LSST science goals (77\% for NEOs). This equates to reducing the undiscovered population of PHAs by additional 26\% (15\% for NEOs), relative to the baseline survey.
\end{abstract}


\section{Introduction}

The small-body populations in the Solar System, such as asteroids, trans-Neptunian objects (TNOs)
and comets, are remnants of its early assembly. Collisions in the main asteroid belt between Mars and
Jupiter still occur, and occasionally eject objects on orbits that may place them on a collision course
with Earth. About 20\% of this near-Earth Object (NEO) population pass sufficiently close to Earth orbit such that
orbital perturbations with time scales of a century can lead to intersections and the possibility of collision. These objects that pass within 0.05 AU of Earth's orbit are termed potentially hazardous asteroids (PHAs).
In order to improve the quantitative understanding of this hazard, in December 2005 the U.S. Congress
directed\footnote{National Aeronautics and Space Administration Authorization Act of 2005 (Public Law 109-155), January 4, 2005, Section 321, George E. Brown, Jr. Near-Earth Object Survey Act} NASA to implement a NEO survey that
would catalog 90\% of NEOs with diameters larger than 140 meters by 2020 (known as the George
E. Brown, Jr.\ mandate). It is estimated that there are approximately 30,000 such objects 
\citep{2012ApJ...752..110M, harris15, 2016Natur.530..303G, 2017Icar..284..114S}, with just over 7,500 currently known.
For a compendium of additional information about NEOs and PHAs and an up-to-date summary of
discovery progress, see NASA's CNEO webpage\footnote{\url{https://cneos.jpl.nasa.gov/}}.

The completeness level set by the Congressional mandate could be accomplished with a 10-meter-class
ground-based optical telescope, equipped with a multi-gigapixel camera and a sophisticated and robust data
processing system \citep[see NASA-commissioned reports by ][]{stokes03,shapiro10}. 
The Large Synoptic Survey Telescope\footnote{http://lsst.org} (LSST), currently being
constructed, approaches such a system. A concise LSST system description, discussion of science drivers, and other
information, are available in \cite{LSSToverview}.

The LSST baseline strategy for discovering Solar System
objects is predicated on two observations of the same field per night, spaced by a few tens of minutes, and
a revisit of the same field with another pair of observations within a few days. The main reason for two
observations per night is to help association of observations of the same object from different nights,
as follows. The typical distance between two nearby asteroids on the Ecliptic, at the faint fluxes probed by
LSST, is a few arcminutes (object counts are dominated by main-belt asteroids). Typical asteroid motion
during several days is larger (of the order a degree or more) and thus, without additional information,
detections of individual objects are ``scrambled''. However, with two detections per night, the motion
vector can be estimated. The motion vector makes the linking problem much easier because
positions from one night can be approximately extrapolated to future (or past) nights. The predicted
position's uncertainty is typically of the order of several arcminutes, rather than a degree, which effectively
``de-scrambles'' detections from different nights (for a detailed discussion of this algorithm, see \citealt{kubica07} as well as Appendix~\ref{sec:appMOPS} for a theoretical derivation of expected scalings).

Early simulations of LSST performance presented by \cite{IvezicNEO2007} showed that the 10-year baseline
cadence would result in 75\% completeness for PHAs greater than 140 m (more  precisely, for PHAs with
$H<22$). They also suggested that with additional optimizations of the
observing cadence, LSST could achieve 90\% completeness. An example of such an optimization was discussed
by \cite{LSSToverview} who reported that, to reach 90\% completeness, about 15\% of observing time would
have to be dedicated to NEOs, and the survey would have to run for 12 years. More recently, estimates of
LSST yields have been revisited by \citet{GMS2016} (predicted PHA completeness of 62\% for LSST alone) and \citet{VeresChesley2017neo} (predicted PHA completeness of 65\% for LSST alone).
The latest LSST simulation results, presented in Section~\ref{sec:opsim}, yielded a completeness of $\sim$66\% for
PHAs with $H<22$, using the current 10-year baseline survey. The differences in reported completeness
between these studies are attributable to differences in the simulated NEO populations and other modeling
details (such as an improved understanding of the hardware and updated cadence and sky brightness models).
This is discussed further in \S\ref{sec:other}.

These completeness estimates are based on an implicit assumption that 3 pairs of observations
obtained within a 15-30 day wide window are sufficient to recognize that these observations belong
to the same object, and to estimate its orbital parameters (the same criterion has been used in NASA
studies\footnote{See \url{http://neo.jpl.nasa.gov/neo/report2007.html}}).
This so-called linking of individual detections into plausible orbital tracks
will be performed using a special-purpose code referred to as the Moving Object
Processing System (MOPS).

The capability and effectiveness of LSST for discovering moving objects has been questioned (e.g., \citealt{GMS2016}) on
two grounds:
\begin{itemize}
\item A large number of false detections due to problems with image differencing software may
make linking problem prohibitively hard for MOPS. In particular, this objection is motivated by the experience
from extant surveys, such as Pan-STARRS1 \citep{denneau13} or the Catalina Sky Survey \citep{2003DPS....35.3604L}.
\item Modifications of LSST baseline cadence, including image depth, sky coverage and cadence,
required to reach 90\% completeness level, have not yet been explicitly demonstrated using detailed
operations simulations, and made available to the community.
\end{itemize}
We aim to address these critiques here: the two major questions addressed by our study can be informally
stated as ``Will MOPS work?'' and ``If MOPS works, what fraction of  NEOs will LSST discover?''.

We use a combination of sophisticated simulations and real datasets to address these questions.
The main analysis components presented here include:
\begin{enumerate}
\item Analysis of the performance of prototype LSST image differencing software, with emphasis on the rate and
    properties of false detections (so-called ``false positives''), using DECam imaging data.
\item Analysis of the linking of asteroid detections in the presence of a large number of false detections, using MOPS
         and simulated observations.
\item Analysis of a large number of modified observing cadence simulations, coupled with NEO population
          models, to forecast discovery rates.
\end{enumerate}

In \S2 we provide a brief overview of LSST and its strategy for discovering moving Solar
System objects. We measure the performance of prototype LSST image differencing pipeline
in \S3, and MOPS performance in \S4. Modifications of the baseline cadence designed to
boost NEO/PHA completeness are examined in \S5.

We conclude that i) LSST implementation of MOPS can cope with the anticipated false detection rates
in LSST difference images, and that ii) the NEO discovery performance of the LSST baseline cadence
can be appreciably boosted by adequate modifications of the observing strategy. These findings are summarized and discussed in \S6.

\section{LSST Strategy for Discovering Solar System Objects}
\label{sec:strategy}

We briefly describe the LSST system design and observing strategy, and discuss in more
detail image processing and moving object detection.

\subsection{A Brief Overview of LSST  Design}

LSST will be a large, wide-field ground-based optical telescope system
designed to obtain multiple images covering the sky visible
from Cerro Pach\'{o}n in Northern Chile. With an 8.4m (6.7m effective) 
primary mirror, a 9.6 deg$^2$ field of view, and a 3.2 Gigapixel camera,
LSST will be able to image about 10,000 square
degrees of sky per night, with a fiducial dark-sky, zenith 5$\sigma$ depth
for point sources of $m_5(r)$=24.38 (AB). The system is designed to yield
high image quality (with a median delivered seeing in the $r$ band of
about $0.8\arcsec$) as well as superb astrometric  and photometric
accuracy\footnote{For detailed specifications, please see the LSST
Science Requirements Document, \url{http://ls.st/srd}}. The total survey
area will include $\sim$30,000 deg$^2$ with $\delta<+34.5^\circ$, and
will be imaged multiple times in six bands, $ugrizy$, covering the
wavelength range 320--1050 nm. The project is scheduled to  begin the
regular survey operations at the start of next decade.

LSST will be operated in a fully automated survey mode. About 90\% of the
observing time will be devoted to a deep-wide-fast survey mode which will
uniformly observe a 18,000 deg$^2$ region about 800 times (summed over
all six bands) during the anticipated 10 years of operations, and yield a coadded map
to a depth of $r\sim27.5$. These data will result in catalogs including about
$40$ billion stars and galaxies, that will serve the majority of the
primary science programs. The remaining 10\% of the observing time
will be allocated to special projects such as a very deep and fast
time domain survey\footnote{Informally known as ``Deep Drilling Fields".}.

\subsection{LSST Observing Strategy}

As designed and funded (by the U.S National Science Foundation and
the Department of Energy), LSST is primarily a science-driven mission.
The LSST is designed to achieve goals set by four main science themes:
\begin{enumerate}
\item Probing Dark Energy and Dark Matter;
\item Taking an Inventory of the Solar System;
\item Exploring the Transient Optical Sky;
\item Mapping the Milky Way.
\end{enumerate}
Each of these four themes itself encompasses a variety of analyses, with
varying sensitivity to instrumental and system parameters. These themes
fully exercise the technical capabilities of the system, such as photometric
and astrometric accuracy and image quality.

The current baseline survey strategy is designed to maximize the overall science returns, including
Solar System science, rather than just the completeness of NEO/PHAs brighter than $H=22$ (though the
two goals are highly interrelated). Discovering and linking objects in the Solar System
moving with a wide range of apparent velocities (from several degrees per day for
NEOs to a few arc seconds per day for the most distant TNOs) places strong
constraints on the cadence of observations. The baseline strategy requires closely
spaced pairs of observations, two or preferably three times per lunation. The visit
exposure time is set to 30 seconds to minimize the effects of trailing for the majority of
moving objects. The images are well sampled to enable accurate astrometry, with
anticipated absolute calibration accuracy of at least 0.1 arcsec (based on an early analysis 
of Gaia's Data Release 1, the accuracy of astrometric calibration will probably improve 
by more than an order of magnitude when using upcoming Gaia's dataset; \citealt{Gaia}).
Typical astrometric errors for LSST detections will range from about 50 mas or better
at the signal-to-noise ratio SNR=100 (dominated by systematics), to 150 mas at SNR=5
(dominated by random errors). 

LSST observations can be simulated using the LSST Operations Simulator tool
\citep[OpSim,][]{delgado14}. OpSim runs a survey simulation with user-defined
science-driven proposals, a software model of the telescope and its control
system, and models of weather and other environmental variables. The output of
the simulation is an ``observation pointing history''; a record of times,
pointings, filters, and associated environmental data and telescope
activities throughout the simulated survey.  This history can be examined using
the LSST Metrics Analysis Framework tool \citep[MAF,][]{jones14} to assess the
efficacy of the simulated survey for any particular science goal or
interest\footnote{For examples of such analysis, see \url{http://ls.st/xpr}}. These
tools -- OpSim and MAF, and the sky brightness, throughput and sensitivity modeling --
are part of the LSST simulation effort \citep{LSSTSimsOverview}, which provides high-fidelity
tools to evaluate LSST performance.

\subsubsection{LSST Baseline Survey Simulation}

As the system understanding improves, the baseline survey strategy and the telescope model
are updated, generally on a yearly schedule. The current
reference baseline simulated survey is known as {\it minion\_1016}. It includes 2.4
million visits collected over 10 years, with 85\% of the observing time spent on the
main survey and the rest on various specialized programs. The median number of visits
{\it per night} is 816, with 3,026 observing nights. The median airmass is 1.23 (the
minimum attainable altitude for the LSST telescope is 20 deg.). In the $r$ band, the median
seeing (FWHM) is 0.81 arcsec, and the median $5\sigma$ depth for point sources is 24.16
(using the best current estimate of the sky background and system throughputs and accounting
for the distribution of observing conditions in the baseline survey; dark-sky, zenith observations
match the fiducial $5\sigma$ depth of $m_5(r)$=24.38).

There are a few known problems with this simulation, including twilight sky brightness
estimates that are too bright, the moon avoidance that is not as aggressive as it could be,
and observations that are biased towards west, away from the meridian. The implied impact
of these shortcomings on NEO completeness estimates is a few percent (the performance
of this simulated cadence in the NEO context is discussed in detail in \S5). An improved simulation,
that will presumably rectify these problems, will become available by the end of 2017.

\subsection{Overview of LSST  Data Management and Image Processing}

The images acquired by the LSST Camera will be processed by LSST Data Management
software \citep{juric15} to a) detect and characterize imaged
astrophysical sources and b) detect and characterize temporal changes
in the LSST-observed universe. The results of that processing will be
reduced images, catalogs of detected objects and their measured properties, and
prompt alerts to ``events'' -- changes in astrophysical scenery discovered by differencing
incoming images against older, deeper, images of the sky in the same direction ({\em
templates}). More details about the data products produced by the LSST Data Management system are described in
LSST Document LSE-163 \citep[LSST Data Products
Definition Document,][]{LSE-163}.

LSST will use two methods to identify moving objects:
\begin{enumerate}
\item Detecting trailed motion on the sky: objects trailed by more
  than 2 PSF widths (corresponding to motion faster than about 1
  deg/day) will be easily identifiable as trailed. LSST will detect sources in 
  difference images using standard point-source detection techniques;
  these sources will then be measured by PSF photometry and by fitting a trailed
  point source model \citep[see][Footnote 41]{LSE-163}).
  Two detections, both recognized as trailed, and within 20--60 minutes in a single night
  will be sufficient to identify an object as an NEO candidate,
\item Inter-night linking of pairs of detections from the same night: the Moving Object 
Processing System (MOPS) will recover objects moving too slow to be measurably elongated in a single exposure.
\end{enumerate}

We note that sources detected in difference images \citep[\DIASources in LSST parlance, see][]{LSE-163}
will also include false detections, colloquially known as {\it false positives}.
In addition to false detections due to instrumental artifacts and software glitches,
in this context they will also include detections of true astrophysical transients
(e.g. gamma-ray burst afterglow) that will not be associated with static sources
(e.g. stars and galaxies). Estimates of expected false detection rates are derived
in \S\ref{sec:imDiff}.

\subsection{The Basic Strategy for Linking Detections into Orbits}

The LSST strategy for linking detections into orbits assumes the following main steps:
\begin{enumerate}
\item Detections in difference images (obtained during the same night), that do not
         have a nearby static object (e.g. variable stars) within a small exclusion radius
         (a fraction of an arcsecond, but possibly larger for brighter stars), are linked into tracklets. There will be of the order
         a million tracklets per observing night (see \S\ref{sec:tracklets}).
\item At least three tracklets obtained in a 15-30 day wide window are linked into
         candidate tracks, using kd-trees and pre-filtering steps based on tracklets' positions
         and motion vectors (see \S\ref{sec:tracks}). These pre-filtering steps result in
         about the same number of false tracks as true tracks on the Ecliptic (of the order
         a million), with the completeness depending on population (e.g. main-belts
         asteroids vs. NEOs) and chosen tunable pre-filtering parameters (generally well above 90\%).
\item Candidate tracks are then filtered further (pruned from false tracks) by
        performing an initial orbit determination (IOD) to assess which tracks
        correspond to Keplerian orbits.
         Due to the high astrometric accuracy of the observations (uncertainties of 0.15 arcsec for the faintest objects,
         see \S\ref{sec:astromerrors}), the number of false tracks which 
         could pass IOD-based filtering becomes negligible, as is the incompleteness induced by this 
         step. A detailed discussion of the IOD step, 
         including an analysis of the accuracy of fitted orbital parameters,  will be presented in 
         a future publication; however an independent JPL study by \citet{VeresChesley2017mops} 
          confirms that orbit determination is indeed a reliable filter. This is further discussed in \S\ref{sec:mopsVeresChesleyComparison}.
\end{enumerate}

Incorrectly linked tracks will not be
an issue when identifying real moving objects because IOD will efficiently and reliably filter 
out false tracks due to the high-accuracy
astrometry and well-understood simple Keplerian model predictions. Therefore, the
essential question is whether the number of detections, including false detections, can be 
linked into tracks, and whether the resulting number of tracks, including false tracks, 
passed to the final IOD step can be handled with available computing resources.

These are all critically dependent on the performance of image differencing codes, and the number of false positive detections they may generate. We turn to these in the next section.

\section{Analysis of Image Differencing Performance \label{sec:imDiff}}

LSST will detect motion and flux variability by differencing each incoming image
against a deep template (built by combining multiple images of the same region).
Sources in difference images, called \DIASources, will be detected at a signal-to-noise
ratio (SNR) threshold of $\nu=5$. Up to about 1,000 deg$^{-2}$
astrophysical, real, detections (e.g. variable stars) are expected in LSST image differencing, including
up to about 500 deg$^{-2}$ asteroids on the Ecliptic. In addition to real detections,
there are false detections due to imaging or processing artifacts and 
false detections caused simply by statistical noise
fluctuations in the background. In a typical LSST difference image, the expected
density of false detections due to background fluctuations is about 60 deg$^{-2}$
(see below for details)---much lower than the expected rate of astrophysical
detections (see \S\ref{sec:kaiser} below). 

Historically, surveys have reported detection rates in image differencing that are much
higher, depending on the survey; see \cite{denneau13}; \cite{kessler15}; \cite{goldstein15}.
For example, Pan-STARRS1 (PS1) reported a transient detection rate as high as 8,200 deg$^{-2}$
\citep{denneau13}. For a ``menagerie'' of PS1 artifacts (with memorable names such as
{\it chocolate chip cookies, frisbee, piano, arrowhead, UFO}), see Fig.~17 in \cite{denneau13}.
They reported that ``Many of the false detections are easily explained as internal reflections,
ghosts, or other well-understood image artifacts, ...''. As discussed in \S\ref{sec:tracklets},
such a high false detection rate is at the limit of what could be handled even
with the substantial computing power planned for LSST.

Over the past decade,
the second-generation surveys have learned tremendously from the PS1 experience. There are surveys
running today, such as Dark Energy Survey, which have largely solved the key problems that
PS1 has encountered. Major improvements to hardware include CCDs with significantly fewer
artifacts (e.g. DECam, see below, and LSST) and optical systems designed to minimize ghosting
and internal reflections (e.g. in LSST). Improvements to the software include advanced image
differencing pipelines (e.g., PTFIDE for the Palomar Transient Factory and the Zwicky Transient
Facility) and various machine-learning classifiers for filtering false detections. For example,
\cite{goldstein15} used a Random Forest classifier with the Dark Energy Survey data
and cleaned their sample of transient detections from a raw false:true detection
rate ratio of 13:1 to a filtered rate of 1:3. The resulting false detections are
morphologically much simpler; for example, compare
Fig.~1 in \cite{goldstein15} to Fig.~17 in \cite{denneau13}.

Here we summarize an analysis of image differencing performance based on DECam
data and difference images produced and processed using prototype LSST
software \citep{DMTN-006}. This analysis demonstrates that the false
detection rate anticipated for LSST (without using any machine-learning classifiers for
filtering false detections) is significantly below the threshold for
successful deployment of MOPS, as will be discussed in \S\ref{sec:mops}.

\subsection{LSST Image Differencing Pipeline and Data Processing}

The LSST prototype image differencing and analysis code derives from the
algorithms employed in the HOTPANTS package \citep{becker15}, and was used for surveys such as SuperMACHO
\citep{becker05} and ESSENCE \citep{miknaitis07}. While this software is
functional as-is, it is expected that the ultimate LSST pipeline will include
improved methods for handling observations at high airmass and the effects of
differential chromatic refraction due to the Earth's atmosphere. In
this work, we conservatively assume that the pipeline used for LSST will have the
same performance as the current code.

\subsection{False Detections due to Background Fluctuations \label{sec:kaiser}}

Some false detections are expected simply due to background fluctuations, even
at a high SNR threshhold. The number of
such detections, as a function of the threshold SNR, the number of pixels and
seeing, can be computed using the statistics of Gaussian random fields.
For an image with a Gaussian background noise, convolved with a Gaussian point
spread function (PSF) with width $\sigma_g$ (in pixels), the number of peaks, $N$, above a
given SNR threshold, $\nu$, is given by (N. Kaiser, priv. comm.)
\begin{equation}
N(>\nu)  = \frac{n_{row}*n_{col}}{2^{5/2} \, \pi^{3/2} \, \sigma_g^2} \, \nu \, e^{-\nu^2 /2}
\label{eq-theory}
\end{equation}
where $n_{row}$ and $n_{col}$ are the number of pixel rows and columns in the image.
This expression was verified empirically by LSST data management team using
raytraced image simulations\footnote{See \url{https://github.com/lsst/W13report}}.
For 4k by 4k LSST sensors the pixel size is 0.2 arcsec, and for a nominal seeing
of 0.85 arcsec and $\nu=5$, $N(>\nu) = 59$ deg$^{-2}$.

It is generally not well appreciated just how steep is the dependence of $N(>\nu)$
on $\nu$ due to the exponential term. Changing the threshold from 5 to 5.5
decreases the expected rate by a factor of 12, and the rate increases by a factor
of 9.7 when the threshold is changed from 5 to 4.5. In practice, an empirical estimate
of the background noise is used when computing the SNR for each detected source.
When this estimate is incorrect, e.g. due to reasons discussed below, then the
implied detection threshold is wrong, too. For example, if the noise is underestimated
by only 10\%, the computed SNR will be too large by 10\%, and the adopted
threshold $\nu=5$ will actually correspond to $\nu=4.5$ -- and thus the
sample will include 9.7 times as many false detections due to background
fluctuations! Hence, the noise in difference images has to be estimated to
high accuracy.

\subsubsection{The Impact and Treatment of Correlated Noise}

When the LSST pipeline convolves the science image to match the PSF of the template
image, the per-pixel variance in the image is reduced, and at the same
time correlations between neighboring pixels are introduced. This violates the
assumption made by standard image processing algorithms that each pixel is an
independent draw from a Poisson distribution. The per-pixel noise reduction is
reflected in the variance plane that accompanies each exposure during
processing, but the covariance between pixels is not tracked.
The significance of detections and the uncertainties on source measurements is
then estimated based on this incomplete information provided by the variance
plane, leading to a biased detection threshold.

The magnitude of this effect can be large---using only the per-pixel variance
measurements can result in underestimating the true noise on PSF-size scales by
20\% or more. A detection threshold of $\nu=5$ thus actually corresponds to
$\nu=4$, and it is easy to see using eq.~\ref{eq-theory} that this error results
in an increase in the number of false detections by a factor of $\sim$70!

\begin{figure}
  \centering
  \plotone{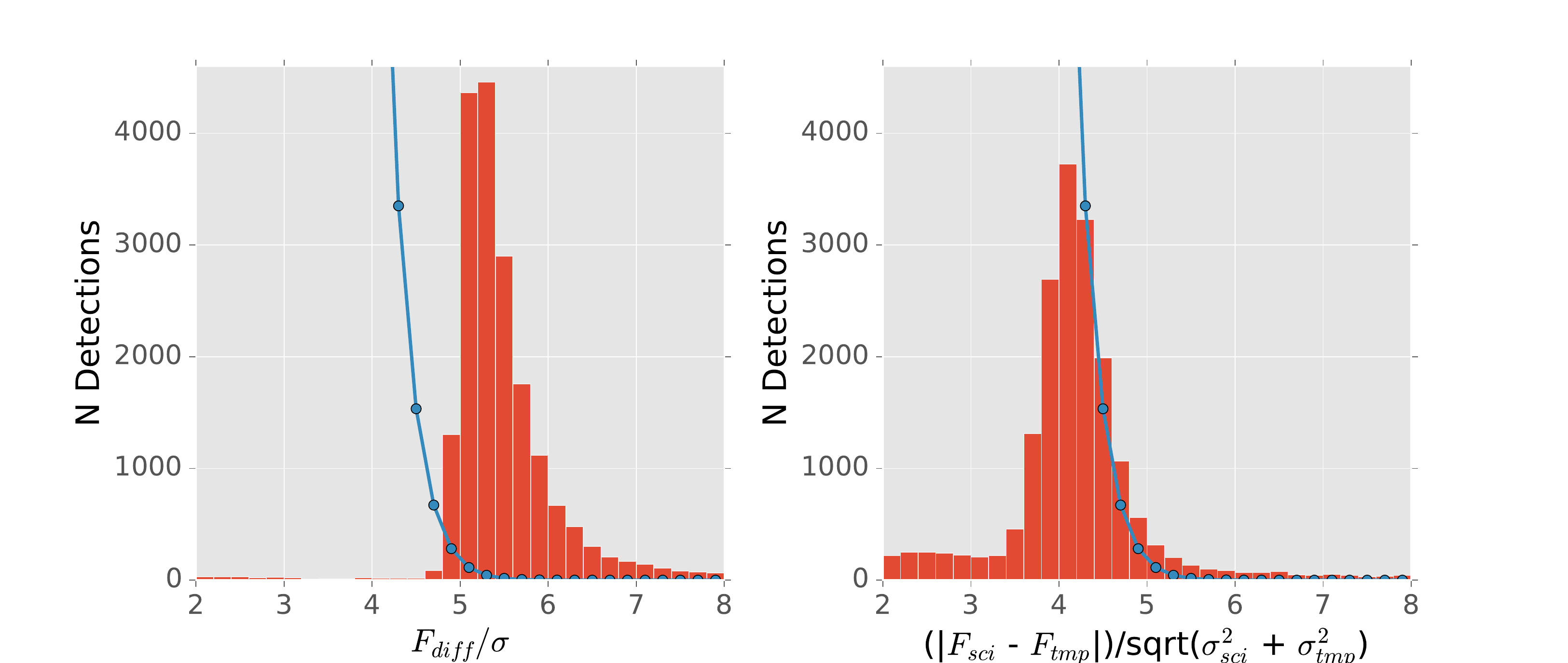}
  \caption{
  Histogram of the reported SNR of sources measured in the difference images,
  using two different SNR estimates: SNR (incorrectly) estimated using the variance plane
  of the difference images (left) and SNR (correctly) estimated from forced photometry on
  the input images (right).  The blue lines indicate the expected SNR distribution
  based on Gaussian background noise. The difference between these two histograms
  illustrates the strong impact of the SNR mis-estimation. Using the correct SNR values,
  the vast majority of putative $>5 \sigma$ detections become $<5 \sigma$ detections
  and can be disregarded.
  }
  \label{fig:snr_comparison}
\end{figure}

A histogram of the number of sources detected in difference images, as a function
of SNR computed using forced photometry measurements, is shown in the right panel
of Figure~\ref{fig:snr_comparison}. The blue line shows the expected counts given
by eq.~\ref{eq-theory}, in good agreement with data. When the SNR is estimated
incorrectly due to correlated noise (left panel), the distribution clearly ramps
up at a much higher SNR value and results in numerous false detections that are
mis-classified as $>5 \sigma$ detections.

Tracking the covariance caused by multiple convolutions is a planned feature for
the LSST software stack, but is not currently implemented. Previous surveys, such as
Pan-STARRS1, have used a small covariance ``pseudo-matrix'', which tracks the
covariance between a small region of neighboring pixels, and then assumed that
this relationship between pixels is constant across an image (Paul Price, priv. comm.).
This method avoids the creation of the full $N_{\rm pixels}$ by $N_{\rm pixels}$
covariance matrix, which is impractically large and mostly empty.

In the interim, for this analysis we have mitigated the problem by utilizing forced photometry of \DIASources
on individual images (that is, before convolution to match their PSFs). This
produces both flux measurements and associated uncertainties which are not
affected by covariance, enabling us to accurately set a SNR threshold
that recognizes and rejects all \DIASources with $\nu < 5$. This mitigation step
will be unnecessary once the image covariance tracking is properly implemented
in the LSST stack. Alternative solutions such as image ``decorrelation''
\citep{DMTN-021}, or the \citet{zackay} image differencing algorithm, would also
alleviate the covariance problem. Tests of these methods in the LSST pipeline are ongoing.

\subsection{Testing the LSST Pipeline with DECam}

The Dark Energy Survey (DES) is an optical/near-infrared survey that aims to probe the
dynamics of the expansion of the universe and the growth of large scale structure by
imaging 5,000 sq. deg. of the southern sky. DECam, the imaging camera developed for
DES, is sufficiently similar to LSST camera to enable an informative study of false detection
rates: DECam includes 62 mosaicked deep-depletion CCDs, with a total pixel count of
520 Mpix over its 3 deg$^2$ field of view, and has a similar filter complement as LSST \citep{DECam}.

The data we use here are a subset of a DECam NEO survey (PI: L. Allen, NOAO) conducted
in the first half of 2013. The data for a given field consist of 40-second exposures separated
by about five minutes. Due to the difference in telescope aperture, these images
are about 1 mag shallower than the 30 second visits by LSST.

Throughout this section, we used data for
five different fields, each with between three and five visits for a total of 15
``science'' visits plus 5 template visits. We've subsequently verified the findings using a much larger superset of 540 visits from the same survey. The enlarged set was bounded by $28^\circ < b < 58^\circ$ and $315^\circ < l < 350^\circ$ (Galactic coordinates), covering a range of object densities better representative of the LSST's "Wide-Fast-Deep" (WFD) survey (except for the depth, which we discuss in \S\ref{sec:imDiffScale}). Using this set, we find false detection rates broadly consistent with those from the smaller subset: the additionally processed visits actually have {\em fewer} false detections by about 30\% (on average).

From the NOAO archive we obtained images that had already had instrumental
signature removal applied by the DECam Community Pipeline. Each image was
processed through the initial LSST pipeline for background subtraction, PSF
determination, source detection and measurement (collectively
termed ``processCcd'' in the LSST pipeline). For each field, we arbitrarily
selected one of the visits to serve as the ``template'' exposure, against which
the other visits in the field are differenced. Sources are then detected in the
difference image to produce \DIASources, and forced photometry performed in both
the original ``science'' and ``template'' exposures at the position of any
\DIASource.

In LSST operations, coadded prior exposures will be used
as templates for image differencing rather than single visits, which will reduce
the noise in template images. In this study, our use of single visits instead of
coadded templates implies that some moving objects or transients will appear as
negative sources in the difference images. We simply disregard these sources
since our goal is to mimic LSST operations rather than discover all possible
transients in this dataset.


\subsubsection{The Transient and False Detection Rates in DECam Images}

Using the $>5\sigma$ cut based on SNR estimated using forced photometry,
the average number of positive \DIASources is $\sim 1000$ deg$^{-2}$,
with some fields having as few as $500$ deg$^{-2}$. A large fraction of
these detections are the result of stars that have been poorly-subtracted and
left significant residuals in the difference image. It is a common problem for
subtracted stars to exhibit ``ringing'' with both positive and negative
excursions, and these images are no exception. Because the focus of this work
is on detecting moving objects rather than variable stars or transients, we have
not attempted to correct these subtraction artifacts. Instead, we simply exclude
difference image detections where there is significant ($>15 \sigma$) flux in both the
science and template images at the position of the \DIASource---that is, we exclude all \DIASources that
overlap with a static source (of course, some may be truly variable sources).
The area lost due to this masking is less than $1\%$ of the total sky. Again, this
is not the intended behavior of LSST during production, but instead a temporary
expedient we can use for conservatively estimating the system's performance.
One could view this step as a ``poor man's'' machine learning step.

After excluding all \DIASources associated with stationary objects, the
remaining candidate moving object detections number on average $\sim 350$
deg$^{-2}$. This sample includes asteroids, false positives, and possibly some
true astrophysical transients that are not associated with stationary objects
(gamma-ray burst afterglows, very faint variable stars with sharp light
curve maxima, etc.). To improve the estimate of the fraction of these remaining
objects that are false, we visually classified one focal plane of detections either
as obvious imaging artifacts, obvious PSF-like detections, or unidentifiable
detections. Approximately 25\% of the reported detections were clearly some
sort of uncorrected artifact (we did not pursue the cause of individual artifacts),
25\% appeared to be acceptable PSF-like features, and the remaining 50\% were
ambiguous or had too low of signal to noise to be able to classify.
Therefore, a conservative upper limit on the fraction of false detections is 75\%,
assuming the 25\% of detections which had acceptable PSF-like features were real objects,
corresponding to a rate of 263 deg$^{-2}$. Given the size of DECam pixels
(0.263 arcsec) and typical seeing of about 1.1 arcsec ($\sigma_g$ = 1.8 pix),
the expected rate due to background fluctuations is 33 deg$^{-2}$, leaving
a rate of 230 deg$^{-2}$ as ``unexplained'' false detections.

The SNR distribution of this sample is proportional to 1/SNR$^{2.5}$, which
is similar to distributions expected for astrophysical objects. This fact implies
that the sample might be dominated by true astrophysical transients and
asteroids; nevertheless, we adopt the above conservative upper limit of 75\%.

\subsubsection{Scaling DECam Results to LSST Performance \label{sec:imDiffScale}}

The LSST false detection rate due to background fluctuations will be about twice
as large as for DECam because of LSST's smaller pixels and better seeing. The scaling
with pixel size and seeing for ``unexplained'' false detections is not obvious
because their cause is unknown. For example, if they are a pixel-induced effect,
their rate should be scaled up by the square of the ratio of angular pixel sizes, or
a factor of 1.72. If they are instead dominated by true astrophysical transients, 
they should not be scaled at all. Since our dataset contains an unknown 
mixture of false detections from these two types of scalings, in addition to a 
significant number of true detections, we cannot derive a precise scaling for the 
false detection rate. Instead, we adopt a {\it conservative} option by assuming
that {\it all of our detections are false} and behave like pixel-induced effects, 
which scales up the DECam rate for ``unexplained'' false detections to 396 deg$^{-2}$. 
In addition, there will be 60 deg$^{-2}$ false detections due to background fluctuations 
(eq.~\ref{eq-theory}, referenced to the median seeing of 0.85 arcsec).

The total false detection rate of $\sim450$ deg$^{-2}$ anticipated for LSST is thus
comparable to the rate of astrophysical transients. Again, this estimate of the false
detection rate is conservative and it would not be very surprising if it turns out
to be much smaller.

In addition to the uncertainties mentioned above, it is hard to precisely account for 
the fact that LSST images will be about one magnitude deeper than the analyzed DECam 
images. The contribution to false detections from background fluctuations 
(60 deg$^{-2}$) should not be changed because it depends on SNR, not the
specific magnitude limit. If all the remaining
false detections are due to pixel-induced effects, they are also dependent on SNR 
(i.e. counts) and not on magnitude. In this scenario, the number of false detections
would not vary even though LSST images would be deeper. If instead all remaining 
false detections are astrophysical in nature, the DECam rate (230 deg$^{-2}$) should 
{\it not} be multiplied by 1.72 for pixel-scaling, and instead should be corrected for the 
difference in image depth. Since the observed differential
\DIASource distribution scales approximately as SNR$^{-2.5}$, one magnitude of 
depth increases the sample size by a factor of about 4. More precisely, we know
that one third of the ``unexplained'' false detections are likely pixel-induced effects
(the 25\% which were clearly some uncorrected artifact above),
and no more than two thirds can be astrophysical, so the scaled rate expected
for LSST would be (1/3*1.72 + 2/3*4)*230 + 60 = 805 deg$^{-2}$, or about 
a factor of 1.8 higher than the adopted rate of $\sim450$ deg$^{-2}$. We emphasize
that this estimate corresponds to the worst case scenario that is rather unlikely
to be correct.

\subsubsection{Spatially Correlated Transients}

\begin{figure}
  \centering
  \plotone{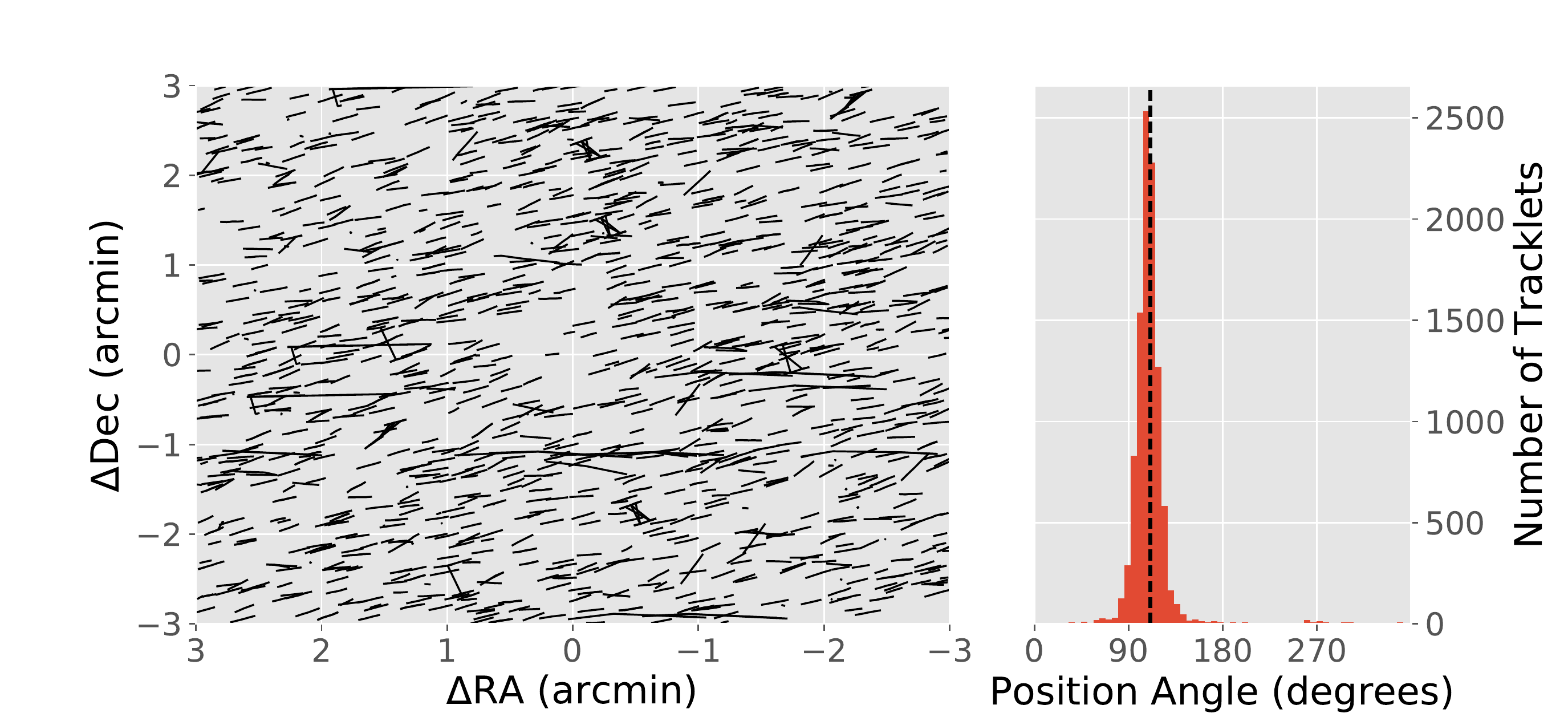}
  \caption{
   ``Stacked'' tracklets of three or more detections surrounding bright stars are shown in the left panel
  as lines. $\Delta \textrm{Dec}$ and $\Delta \textrm{RA}$ are the tracklet coordinates relative to a nearby
  bright star, such that any correlated features due to, e.g., bleed trails or
  diffraction spikes, will show a grouping of lines with similar positions and
  orientations. No such features are seen and the vast majority of
  tracklets are consistent with real moving objects (and on average aligned with
  the Ecliptic). The ``hole'' in center of the plot comes from the rejection of \DIASources
  in regions with significant flux in the template image. The right panel shows
  the position angle of the tracklets (degrees East of North), while the dashed
  vertical line indicates the angle that would be seen if the tracklets were on
  lines of constant Ecliptic latitude. Again the vast majority appear to be
  aligned with the Ecliptic.
  }
  \label{fig:stacked_tracklets}
\end{figure}

We have checked for correlations in tracklets around bright stars, which
might arise if either optical or processing artifacts exhibited some preferred
alignment. Such an alignment may create false tracklets at a rate greater than
an overdensity of uncorrelated detections would. To investigate these
correlations we generated tracklets using the current prototype version of MOPS.
For this test tracklets are required to have three or more detections (out of five visits
at each pointing) that align with velocities less than $2\deg$/day. For each
star brighter than 14th magnitude in the UCAC4 catalog \citep{UCAC4}, we
identified all tracklets within 4 arcminutes of the star, and ``stacked'' the
tracklets surrounding each star onto a single plot. The resulting tracklet
distribution can be seen in Figure~\ref{fig:stacked_tracklets}, where each black
line corresponds to a single tracklet. If, for example, there was an excess of
tracklets along CCD bleed trails from bright stars, these would appear as a set
of lines along the $\pm\Delta \textrm{RA}$ direction, or a peak at $90\deg$ or
$270\deg$ in the histogram (right panel of Figure~\ref{fig:stacked_tracklets}).
We do not see any such correlated detections after inspecting approximately
5,000 tracklets (the number in the plot is limited for legibility).

We also investigated whether \DIASources from multiple visits are correlated
in pixel coordinates, which might arise from uncorrected detector anomalies.
We analyzed a 4-visit subset of the 15 science visits used in the rest of this
section, and for each \DIASource computed the distance in pixels to any neighboring
\DIASources, even if they appeared on different visits or different fields
(similar to the 2-point correlation function). From this we identified  24
\DIASources that were located within two pixels of another \DIASource. We did
not find any correlation at larger radii. Visual inspection shows that many are
near parts of an image where a defect (such as a cosmic ray, bad column, bleed
trail, etc) had been interpolated over, though for some the cause is unclear.
The implied density of correlated \DIASources is about 2.3 deg$^{-2}$, rendering
this effect relatively unimportant.

We note that the number and implied sky density of asteroids bright enough to produce
scattered light and diffraction spike artifacts, which would appear to move at solar system
rates, is at least two orders of magnitude smaller than the sky density of asteroids down
to LSST depth on the Ecliptic. Therefore, such artifacts will be unimportant as a contributor
to false tracks.

\section{Analysis of Moving Object Processing System Performance \label{sec:mops}}

The linking of individual detections from difference images into plausible orbital tracks will be performed using
a special-purpose code referred to as the Moving Object Processing System (MOPS). There are several slightly modified
versions of MOPS in use by various projects; the original version was developed collaboratively by Pan-STARRS
and LSST, and is described in \cite{denneau13}. MOPS employs a two-step processing: first pairs of detections
from a given night are connected into {\it tracklets}, and then at least three tracklets are associated into a
candidate {\it track}. Realistic MOPS simulations show high linking efficiency ($>$99\%; \citealt{denneau13})
across all classes of Solar System objects. The core algorithmic components of MOPS are {\it findTracklets} and
{\it linkTracklets} kd-tree algorithms by \citet{kubica07}. {\it findTracklets} links \DIASources from a single
night to produce {\it tracklets}, and {\it linkTracklets} links tracklets from at least three nights to produce candidate
{\it tracks} (assuming quadratic motion in each coordinate; the LSST version also accounts for topocentric
corrections). Candidate tracks produced by MOPS are then filtered using initial orbital determination (IOD) step,
which is executed using a stand-alone code (e.g. OrbFit, \citealt{milani08}; OpenOrb, \citealt{OpenOrb2009}).

Given the empirically estimated false detection rates expected for LSST, discussed in the preceeding section,
in this section we show that MOPS performance is already adequate - MOPS requires significantly less
computing capacity than planned for other LSST data processing needs. In addition to reporting the results of
numerical experiments with MOPS, we also analyze them using analytic and semi-analytic results for the
rates of false tracklets and false tracks.

\subsection{A Summary of LSST tests of MOPS}

As a part of the Final Design Review preparations, the LSST team has developed an enhanced prototype
implementation of MOPS and analyzed its behavior. Here we summarize the main results of that work;
a detailed internal technical note has been made public in \cite{LDM-156}.

Simulated \DIASources were based on a Solar System model by \citet{Grav2011}.
The model includes about 11 million objects; about 9 million are main-belt asteroids. Observations span
30 days and were selected from a simulated baseline cadence (at that time, the baseline simulation was
OpSim3.61, which in this context is statistically the same as the current baseline cadence, {\it minion\_1016}).
The number of tracklets and tracks, the runtime, and the memory usage were studied as functions of
the false positive detection rate. The rate was varied from none to four times the asteroid detection rate
(100 deg$^{-2}$).  The highest rate corresponds to the expected false positive detection rate for LSST
($\rho_{FP} =  400$ deg$^{-2}$).

Tests were run with 16 threads on single 16 CPU node on Gordon cluster at San Diego Supercomputing
Center (in 2011). Due to computational constraints, a $v < 0.5$ deg/day velocity limit for pairing detections
into tracklets was imposed. For similar reasons, the filters that were imposed on track fitting were not
optimized, artificially reducing the yield. As we now understand the algorithmic scalings much better
(see Appendix~\ref{sec:appMOPS}), it is clear that these unoptimized filters have no major impact on the
simulation results and derived conclusions.

As expected, the addition of false detections increases the number of tracklets and tracks,
the runtime, and the memory usage. For the 4:1 false:true detection rate ratio, compared to case with
no false detections, the number of tracklets increases by about
a factor of 10, the number of tracks by about 50\%, and runtime increases by about a factor of 3.
For the 4:1 false:true detection rate ratio, the runtime with 16 CPUs is 33 hours, with maximum memory
usage of about 80 GB.

\begin{figure}[t!]
\centering
\vskip -0.3in
\includegraphics[width=0.95\textwidth]{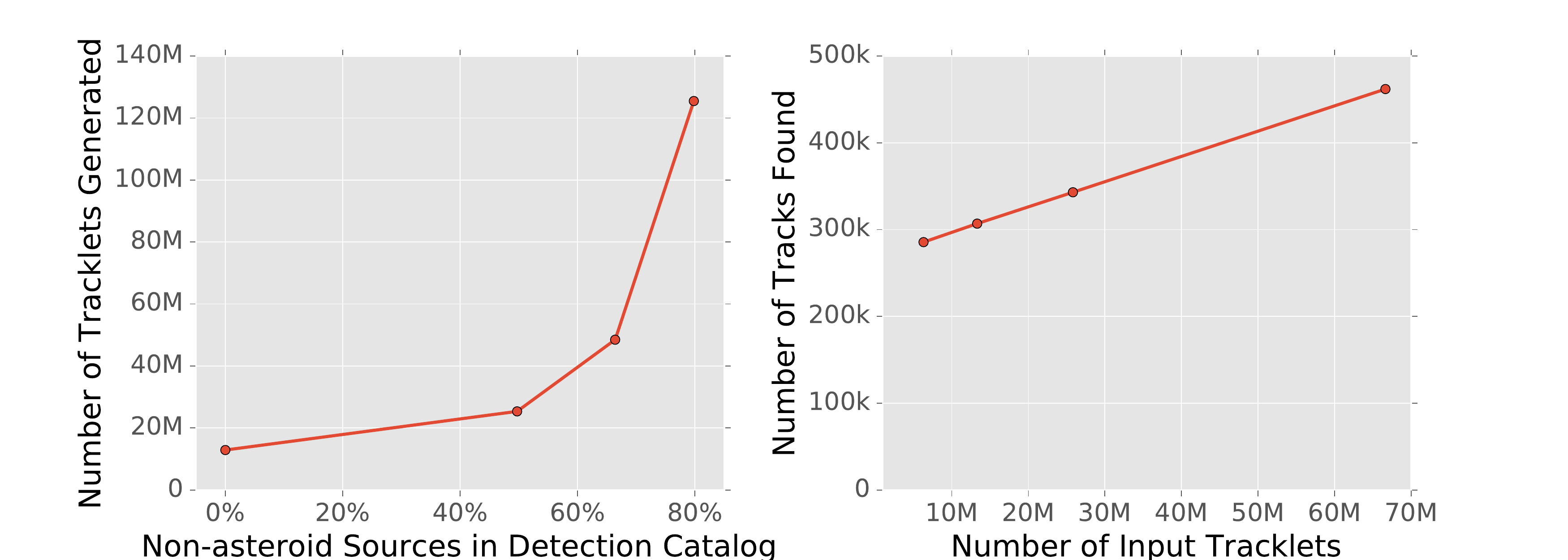}
\caption{A summary of MOPS tests for the dependence of the number of tracklets (left)
and tracks (right) on the false detection rate. As the rate of false detections
increases from none to four times the asteroid detection rate, the number of tracklets
increases by about an order of magnitude. At the same time, the number of candidate
tracks increases by only about 50\%.
\label{fig:MOPStests}}
\end{figure}

\subsection{Understanding MOPS Performance}

The rather slow increase of the number of tracks with false positive detection rate (a 50\% increase
although the number of tracklets increased by a factor of 10) may seem surprising. We have
developed analytic and semi-analytic analysis to better understand the scaling of the number of
tracklets and tracks with false detection rate and other relevant parameters. Details of this
analysis are provided in Appendix~\ref{sec:appMOPS}. Here we briefly discuss the main results.

The increase of the number of tracklets with the false detection rate,
$\rho_{FP}$, shown in left panel in Figure~\ref{fig:MOPStests}, is well
described by eq.~\ref{eq:NttFalse}. In particular, the number of tracklets
approximately increases proportionally to $(C_1 + C_2\rho_{FP}^2)$, where $C_1$
and $C_2$ do not depend on $\rho_{FP}$. As both the full analytic result and the
simulations show, false tracklets quickly outnumber true tracklets even at low
false detection rates, resulting in the observed $\rho_{FP}^2$ behavior.

While the number of tracklets is dominated by the false detections, in the
baseline LSST cadence and the nominal noise assumptions under which the MOPS
simulations were run ($\rho_{FP} \leq 500\,\rm{deg}^{-2}$), the number of
tracks is not dominated by spurious detections---instead it is dominated by true
tracks and mislinkages between true objects. This is due to the fundamental
feature of MOPS: the 4-dimensional space of tracks (two coordinates and two
velocity vector components) is sparse at up to moderate levels of contamination,
and at the tested noise levels false tracklets are effectively filtered out. This
behavior accounts for the slow growth in tracks in the right panel of
Figure~\ref{fig:MOPStests}.

As we evaluate the impact of different survey parameters, we can assess the
number of tracks that would be generated (and thus require IOD processing) using
the analytic results developed in Appendix~\ref{sec:appMOPS}. For a given window
width and false detection density, the number of false tracks per search window
that would arise from false detections is given by
\begin{equation}
\label{eq:falsetracks2}
   N^{falsetracks} = 4.5 \times 10^6 \, \left( {N_w \over 30 \, {\rm day} } \right)^{8} \left( {\rho_{FP} \over 400 \, {\rm deg}^{-2} }\right)^{3.7}.
\end{equation}
This expression is valid around fiducial values and assumes $\rho_{ast}=100$ deg$^{-2}$.
The number of true tracks is of the order 10$^6$; therefore, with the baseline
window $N_W=15$ the contribution of false detections is small, while in the
enhanced NEO cadences with $N_W=30$ the contribution is only a factor of a few
times the number of true tracks.

\subsection{Required Computing Resources for MOPS and IOD Processing}

Given the modest computing resources used in MOPS tests described above, the runtime and memory
usage results bode well for LSST processing. Assuming a 1000-core machine dedicated to LSST moving
object processing (corresponding to about 1\% of the anticipated total LSST compute power), MOPS runtime for producing
candidate tracks should not exceed an hour, assuming sufficient parallelization can be achieved.

The IOD step can also be handled with anticipated resources and is trivially parallelizable. The number 
of available IOD computations for a compute system with $N_{core}$ cores and allocated runtime $T_{runtime}$ 
can be expressed as
\begin{equation}
  N_{IOD} = 3.6\times10^8 \left({ 0.1\,{\rm sec} \over T_{IOD}}\right) \,
                                         \left({ T_{runtime}  \over 10\,{\rm hr} }\right) \,
                                         \left({ N_{core}  \over 1000}\right).
\end{equation}
where $T_{IOD}$ is the time it takes to perform one IOD computation on a single core. 

To get a handle on a realistic estimate of $T_{IOD}$, we benchmark an implementation provided
by the \FindOrb code\footnote{\FindOrb source code can be found at \url{https://github.com/Bill-Gray/find_orb/}} by Bill Gray.
\FindOrb is an open source orbit determination software written in C++.
It implements several IOD methods (Gauss, Herget, and V\"{a}is\"{a}l\"{a}),
whose precision are representative of other analogous packages in the field.

For benchmarking purposes, we used \FindOrb to fit an orbit of an 
$a = 2.46, e = 0.57$ asteroid given six observations (a pair each night) 
spanning a 16-day arc. The measurement is performed on a single 3.1 GHz core. 
We find that the IOD using the Gauss method takes only 0.3ms to compute, a negligible fraction of 
our 100ms computational budget. Running an afterburner with Herget's method
to obtain further differential improvements takes 26ms, still comfortably 
within our computational budget.

These numbers leave a significant margin relative to the fiducial value of 100ms adopted here.
They are consistent with anecdotal estimates of JPL's NEO group (not more than 50ms; S. Chesley, priv. comm.). Given that the expected number of candidate tracks to filter using
IOD is well below $10^7$, it should therefore be possible to accomplish the IOD step in well under an hour.
Alternatively, it is plausible that a 100-core machine might be sufficient for LSST moving object
processing (assuming no engineering safety margin).

\subsection{Comparison to \cite{VeresChesley2017mops}\label{sec:mopsVeresChesleyComparison}}

In parallel to the work presented here, \cite{VeresChesley2017mops} have conducted a coordinated but independent evaluation of the linking capabilities using the Pan-STARRS1 variant of MOPS \citep{denneau13}. They perform linking on a three month long simulated LSST dataset and achieve:

\begin{itemize}
	\item 93.6\% NEO linking efficiency for $H < 22$,
	\item 96\% correct linkages in the derived NEO catalog, with
	\item the large majority of false linkages stemming from main belt confusion (that would not arise given a longer simulation), and
	\item less than 0.1\% of orbits due to false detections.
\end{itemize}

\cite{VeresChesley2017mops} findings are fully consistent with the results obtained here using the LSST implementation of MOPS. They show that the linking method underlying both implementations is valid and robust. Furthermore, they confirm that IOD is a sufficient filter for admissible tracks.

Adding our empirical measurement of the performance of LSST image differencing pipelines and the performance of LSST's implementation of MOPS, the two studies demonstrate that LSST's approach to asteroid discovery is on firm footing.

\section{LSST Observing Cadence Optimization to Enhance PHA Completeness \label{sec:opsim}}

LSST takes observations in pairs of visits separated by 20 to 60 minutes. This cadence -- validated in the previous sections -- allows it to increase the sky coverage and improve the survey efficiency relative to taking triplets or quads of visits in each night. There are multiple approaches to covering the sky with pairs of visits. In this section we evaluate the effects of varying the LSST observing strategy and the resulting PHA completeness.

This evaluation is carried out using a combination of the LSST Operations Simulator (OpSim) and the LSST Metrics
Analysis Framework (MAF).
The LSST Metrics Analysis Framework (MAF) is a user-oriented, python package for evaluating the pointing history
from these simulated surveys in light of particular science goals or interests. The various metrics coded in the
MAF framework can be calculated for any given simulated survey and compared as simulation parameters are changed
in OpSim. This permits a thorough investigation of the trades between different observing strategies, in terms of the
effect on multiple science goals, including the PHA completeness.  We first describe the basic steps in our simulations,
then describe the baseline and modified LSST simulated surveys, and then discuss our results.

\subsection{Simulations of LSST Asteroid Discoveries}

The basic components of our end-to-end simulation of asteroid discovery, described in detail below, include
\begin{enumerate}
\item {\it NEO Population Modeling.} Orbital parameters are used to generate asteroid positions during the
simulated survey duration for a simulated or properly debiased extant NEO population. The population needs
to adequately sample color, size and other properties. A database of such positions evaluated with an adequate
time step  is available as an input to MAF.
\item {\it Survey Cadence Modeling.} A series of LSST pointings with instrumental metadata and observing conditions
is generated by OpSim. In addition to boresight positions, the camera orientation and selected filter, available
metadata enable the computation of instrumental sensitivity (limiting magnitudes).
\item {\it Asteroid Optical Flux Modeling.}  Optical flux from an arbitrary asteroid needs to be computed
as a function of the positions of the Sun, the asteroid and Earth, and asteroid physical properties (e.g., size
and color). This model is implemented in MAF.
\item {\it Source Detection Modeling.} Given the instrument model, observing conditions and asteroid flux,
the signal-to-noise ratio is estimated and used to compute detection probability. This model is implemented
in MAF.
\item {\it Detection Linking Modeling.}  Instead of running MOPS, a model that emulates MOPS
performance is used to significantly speed up the computations. This is equivalent to assuming that an object is discovered 
if a given pattern of observations for an object is achieved. This model is implemented
in MAF.
\item {\it Completeness Estimation.} Given a list of ``discovered objects'', and the input population,
the completeness is estimated as a function of asteroid properties (e.g. size) and various other parameters
(e.g. observing strategy). This model is implemented in MAF.
\end{enumerate}

We proceed to describe these models in more detail, and then discuss the baseline and several modified LSST
surveys, and the corresponding PHA completeness estimates.

\subsubsection{NEO Population Modeling \label{sec:MAFdetails}}

\begin{figure}[t!]
\centering
\includegraphics[width=0.49\textwidth]{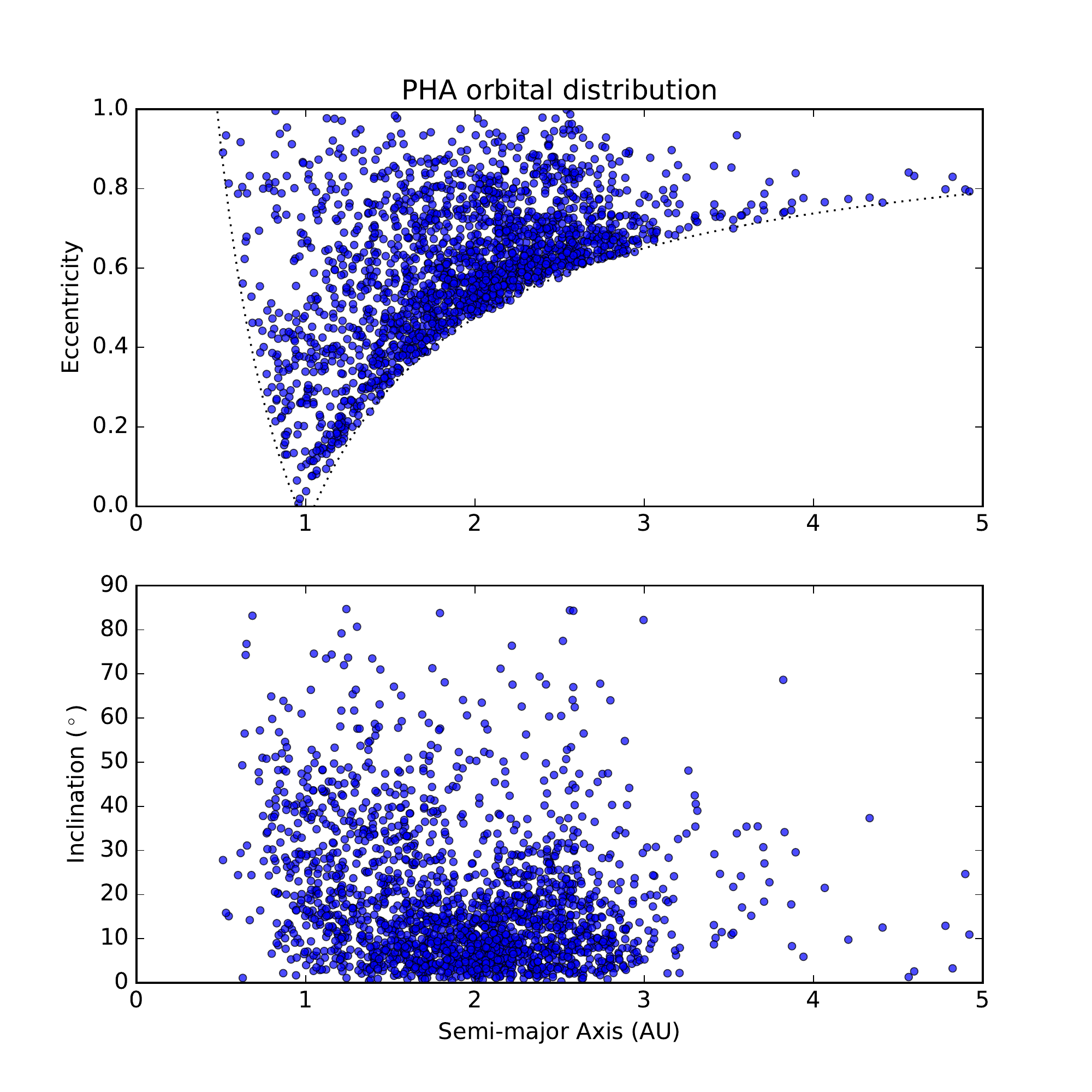}
\includegraphics[width=0.49\textwidth]{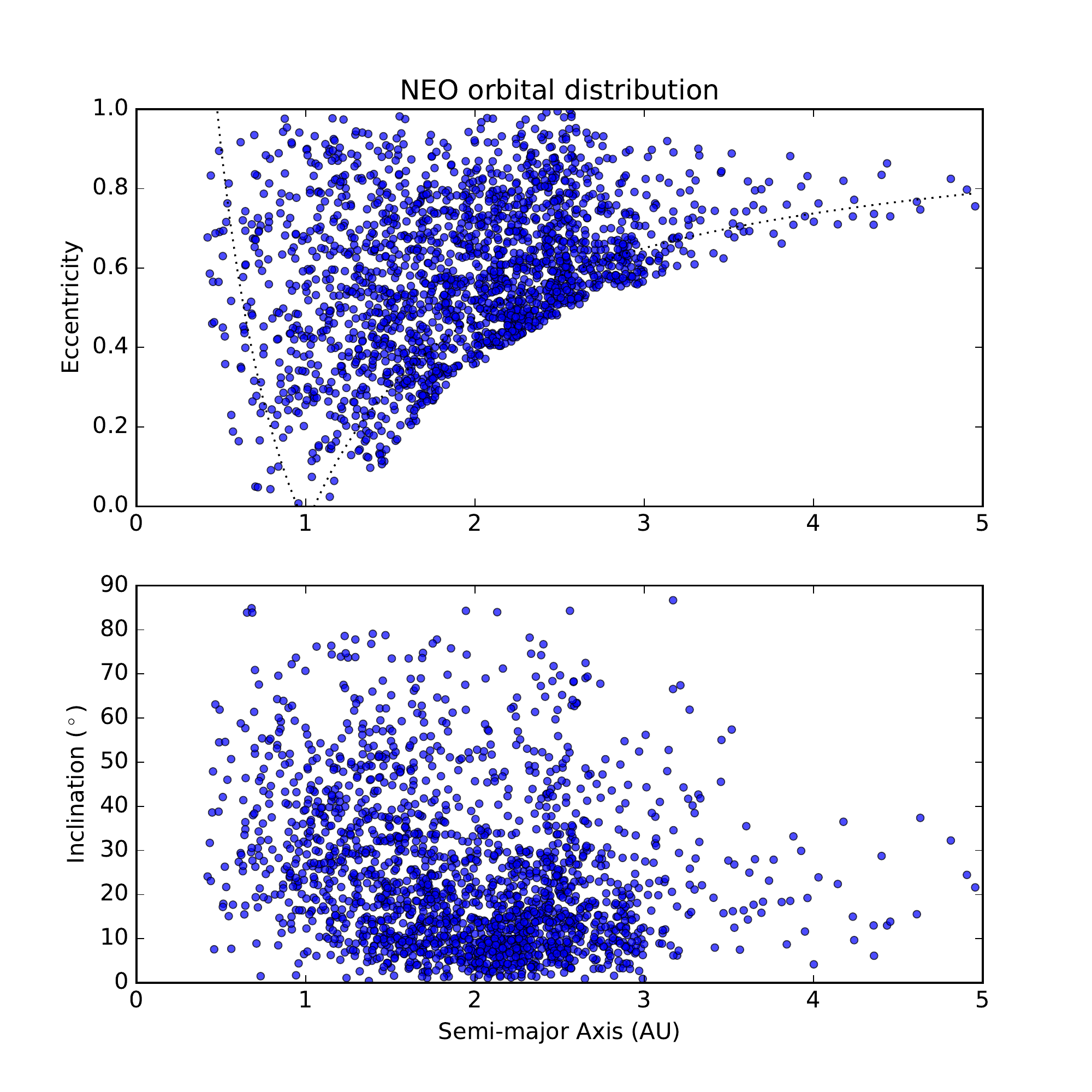}
\vskip -0.2in
\caption{The eccentricity and inclination distributions, as a function of semi-major axis, of the PHAs (left) and NEOs (right) used in this analysis. Both populations were randomly sampled from the S3M model \citep{Grav2011}, a synthetic solar system model based on the \cite{Bottke2002} NEO orbital distribution. NEOs are defined as objects with $q<1.3$~AU; PHAs are defined as having a Minimum Orbit Intersection Distance (MOID) with Earth of less than 0.05~AU (implying $q\le1.05$~AU) and having $H\le22$.  The dotted lines correspond to q = 1.05~AU and Q = 0.95~AU, the approximate limits where MOID$\le 0.05$~AU. \label{fig:PHA_orbits}}
\end{figure}

We use random samples from the synthetic solar system model (S3M) presented in \cite{Grav2011} in order to model completeness for NEOs and PHAs. We have chosen a sample of 2000 NEOs from the \cite{Grav2011} NEO population, which is based on the \cite{Bottke2002} model. We chose a separate sample of 2000 PHAs from the same model, by choosing NEOs with a MOID $\le 0.05$~AU. The PHA population is useful for evaluating PHA completeness directly; the NEO population is useful for comparison to other survey evaluations. A plot of the $a$, $e$, $i$ distributions for these PHAs and NEOs is shown in Figure~\ref{fig:PHA_orbits}.

With this small set of orbits, we then assume that the $H$ magnitude distribution is independent of the orbital distribution.
 For most small body populations, including the PHA population larger than 140~m in diameter, this is approximately true. 
 Assuming an independent distribution, each orbit can be ``cloned'' from the fiducial $H$ magnitude to a range of values 
 covering the interesting sizes for analysis; this allows the analysis to use a large number of objects at each $H$ value, without 
 requiring extensive resources to generate ephemerides for a much larger set of orbits. We use the small population of 
 2000 NEOs or PHAs and clone them to a range of $H$ magnitudes between $H$=14 and $H$=28, with bins of 0.2 in $H$,
 using $dN/dH = 10^{\alpha\, H}$, with $\alpha=0.33$ \citep{2017Icar..284..114S}. We have verified with a larger simulated set 
 of NEOs that reducing the population to 2000 NEOs does not significantly alter the completeness calculation.

Using the details of the input population, MAF generates the expected observations of each object using the pointing history
from a specific OpSim simulated survey. Ephemerides are generated using OpenOrb \citep{OpenOrb2009} for a closely spaced grid of times (typically every 2 hours), and then interpolated to the exact times of each OpSim pointing.

\subsubsection{Survey Cadence Modeling}

The LSST Operations Simulation \citep[OpSim,][]{delgado14} is a Python software package that generates a realistic pointing history, with the time, filter, location, astronomical conditions, weather conditions, and predicted point-source $5\sigma$ limiting magnitude, for each LSST visit
over the lifetime of the survey. This pointing history is generated using weather data (cloudiness and seeing) from the Cerro Pach\'{o}n site and a high-fidelity model of the telescope itself (including slew and settle time and dome movement, for example), combined with a parameterized set of observing proposals that determine how the scheduling algorithm attempts to gather observations. By configuring OpSim with different parameters for the observing proposals, we can generate a series of simulated surveys which prioritize different science goals. The LSST baseline survey and its modifications designed to enhance the PHA completeness are described in detail
in \S\ref{sec:surveys} below.

\subsubsection{Asteroid Optical Flux Modeling}

Given $H$ magnitude for an object, its apparent magnitude in Johnson's $V$ band can be easily computed
given the positions of the object, the Sun and the observer; we have used the two-parameter H-G magnitude 
system \citep{1989aste.conf...21B}, assuming G=0.15.
Magnitudes, or fluxes, in any other optical and near-IR band (in case of LSST, $u$, $g$, $r$, $i$, $z$, and $y$)
can be computed from $V$ magnitude by specifying a spectrum for each object. We have
assumed that our entire NEO population has the same spectral energy distribution as C-type main-belt asteroids.
The computed color transformations for LSST bandpasses are listed in Table~\ref{tab:sed_colors}. Choosing the
spectral energy distribution of  S-type main-belt asteroids instead results in $<$1\% changes in completeness.
These simulation-based colors were verified using SDSS observations \citep{2001AJ....122.2749I} and analogous
computations with SDSS bandpasses.

\begin{deluxetable}{ccccccc}
\centering
\tablecolumns{7}
\tablecaption{Color transformations from Johnson's $V$ band to LSST bandpasses, for C and S type asteroids. \label{tab:sed_colors}}
\tablewidth{0.7\textwidth}
\tablehead{ Type & $V-u$ & $V-g$ & $V-r$ & $V-i$ & $V-z$ & $V-y$  \\ }
\startdata
C  & -1.53 &  -0.28 &  0.18 &  0.29 &  0.30 & 0.30 \\
S & -1.82 &  -0.37 &  0.26 & 0.46 &  0.40 & 0.41  \\
\enddata
\end{deluxetable}

\subsubsection{Source Detection Modeling}

\begin{figure}[t!]
\centering
\includegraphics[width=0.65\textwidth]{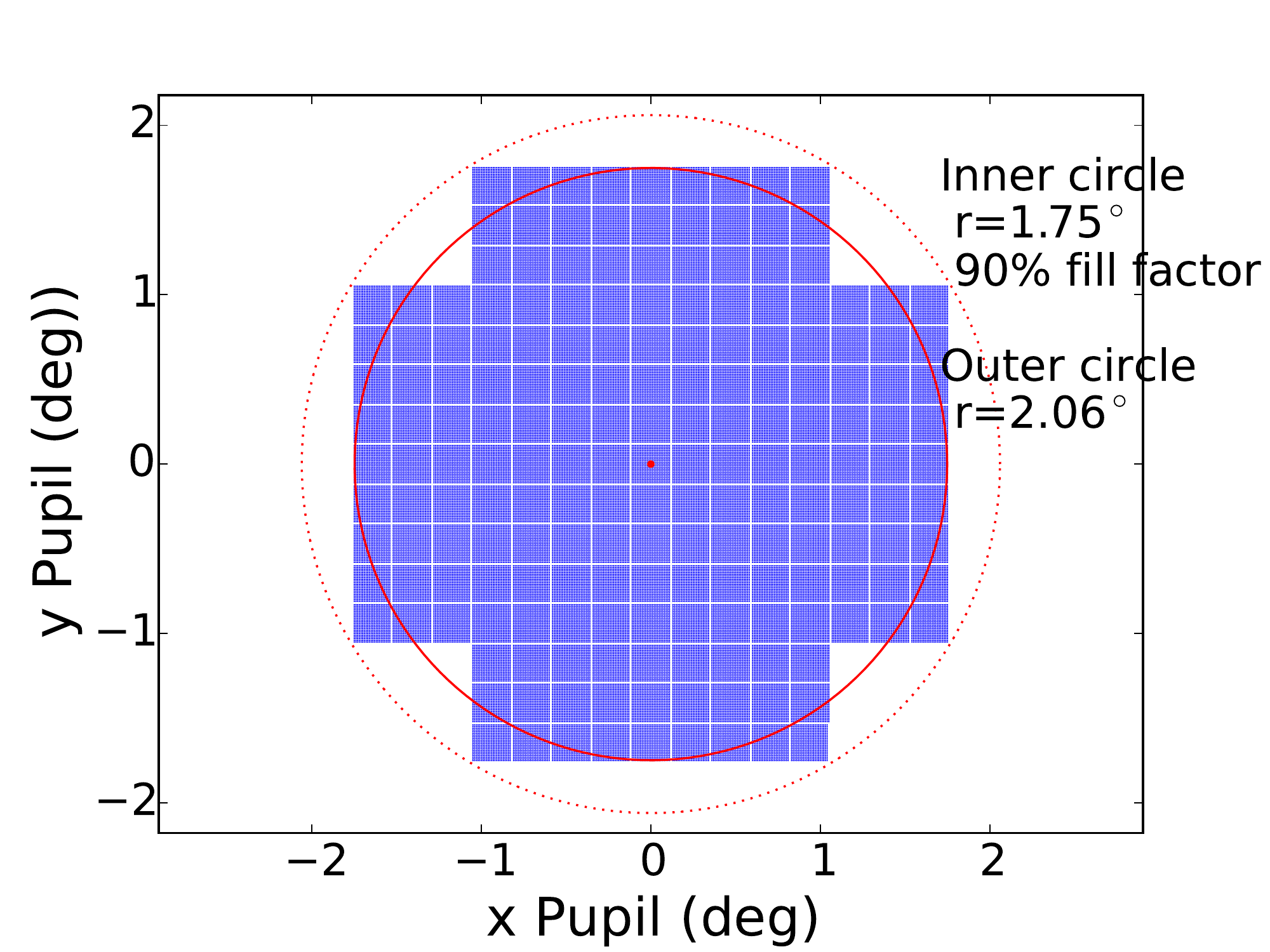}
\caption{Model of the LSST camera footprint, including chipgaps and CCD + raft layout. \label{fig:camera_footprint}}
\end{figure}

If the object is within the LSST field of view, its predicted position, velocity, and apparent $V$ magnitude (calculated from the fiducial $H$ magnitude associated with the orbit) is recorded along with information about the simulated observation itself (such as the seeing, limiting magnitude, filter, and boresight RA/Dec). The full LSST camera footprint (see Figure~\ref{fig:camera_footprint}), including chip gaps, is
used to determine whether an object is within the field of view.

MAF also calculates signal-to-noise (SNR) loss due to trailing for each observation, which is required when evaluating whether
a particular object is detectable in a given observation. Trailing losses occur whenever the movement of an object spreads its photons over a wider area than a simple stellar point spread function (PSF). There are two aspects of trailing loss to consider: SNR losses and detection algorithm losses. The first is the
irreversible degradation in SNR that occurs because the trailed object includes a larger number of background pixels in its footprint, compared to a stationary PSF. The second effect, detection loss, occurs because source detection software is optimized for detecting point sources; a stellar PSF-like matched filter is used when identifying sources that pass above the defined threshold. This filter is non-optimal for trailed objects but losses can be mitigated with improved software ({\it e.g.} detecting to a lower PSF-based SNR threshold and then using a variety of trailed PSF filters to detect sources). When considering whether a source would be detected at a given SNR using typical source detection software, the sum of SNR trailing and detection losses should be used. With an improved
algorithm optimized for trailed sources (implying additional scope for LSST data management), the smaller SNR losses should be
used instead.

\begin{figure}[t!]
\centering
\includegraphics[width=0.85\textwidth]{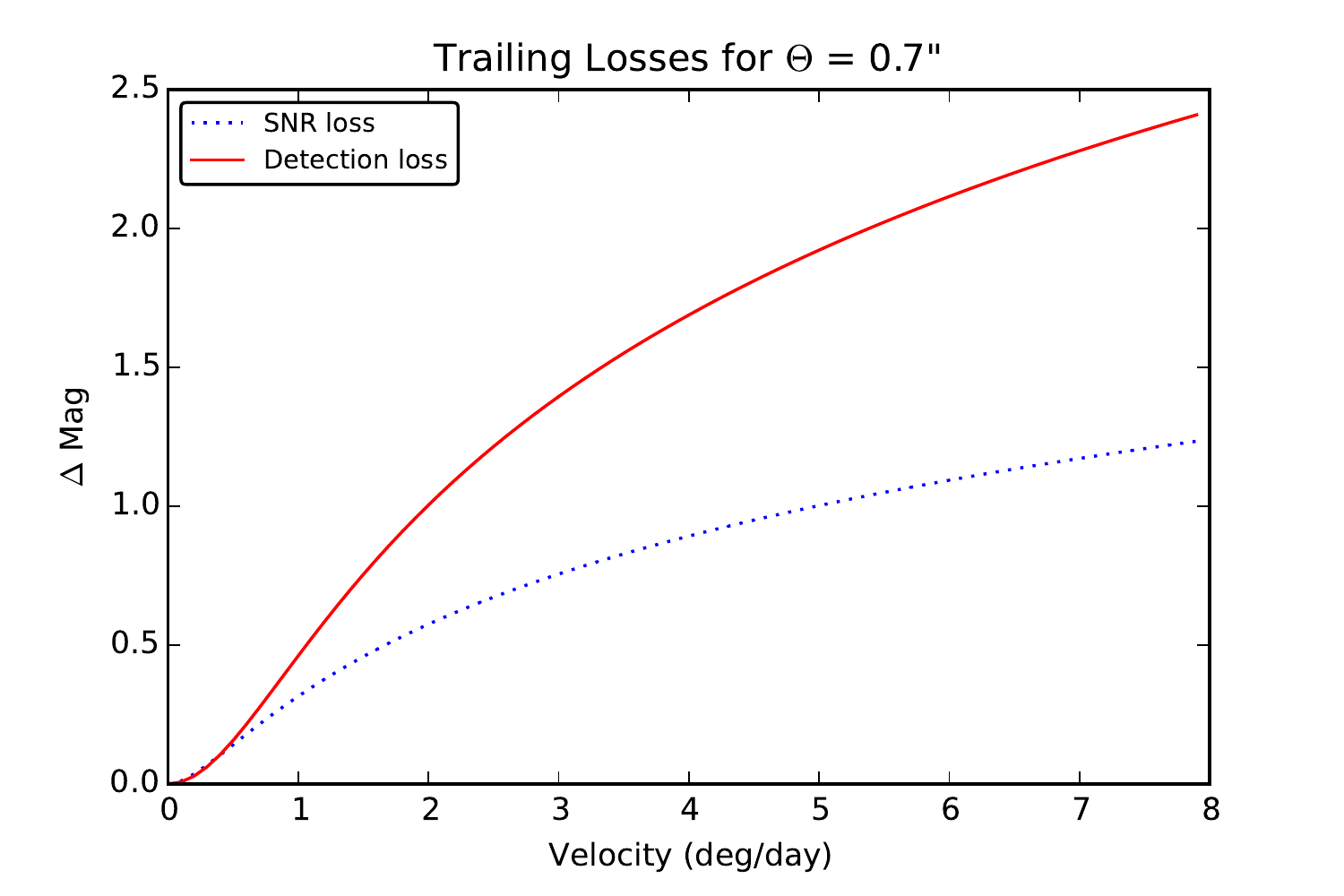}
\caption{Trailing losses for 30 second LSST visits, assuming seeing of
  0.7''. The dotted line shows SNR trailing losses, the solid line
  indicates cumulative losses that also account for non-optimal detection
  algorithm. With software improvements the latter detection losses can be
  mitigated. At the fiducial $v=1$ deg/day, the SNR loss is $\sim$0.3 mag,
  and non-optimal detection algorithm contributes an additional loss of
  $\sim$0.16 mag.
\label{fig:trailinglosses}}
\end{figure}

Our simulations of these effects show that both types of trailing losses can be fit well with the
same functional form:
\begin{eqnarray}
\Delta \, m & = &-1.25 \, log_{10} \left( 1 + \frac{a \, x^2} { 1 + b\,
    x} \right) \\
x & = & \frac{v \, T_{exp}} {24 \, \theta}
\end{eqnarray}
where $v$ is the velocity (in deg/day), $T_{exp}$ is the exposure time (in seconds), and $\theta$ is the FWHM (in arcseconds). For trailing SNR losses, we find $a = 0.67$ and $b = 1.16$; for the cumulative loss, that includes both SNR and detection losses,
we find $a=0.42$ and $b=0$. An illustration of the magnitude of these trailing loss effects for 0.7 arcsec seeing is given in Figure~\ref{fig:trailinglosses}.

We calculate the probability of detecting a particular source given its magnitude $m$
and the $5\sigma$ limiting magnitude $m_5$ (after accounting for trailing losses) using a logistic function
\begin{eqnarray}
     P & = & \left[ 1 +  {\rm exp}\left(\frac {m -  m_5}{\sigma}\right) \right]^{-1}.
\end{eqnarray}
where $\sigma$=0.12 describes the width of the completeness falloff \citep{2014ApJ...794..120A}. A source is randomly classified
as detected using the probability $P$. We also evaluate more optimistic discovery criteria using only SNR trailing losses
(i.e. without taking detection losses into account), as well as detections to SNR=4 instead of SNR=5.  This detection model assumes a flat $m_5$
value across the focal plane; no vignetting or background variation (or masking) due to bright stars is included. These effects however are small.

\subsubsection{Detection Linking Modeling}

Once we have computed the set of all visits in which a given object was within the field of view and detected, we locate subsets of these visits that match our target discovery criteria. These criteria generally consist of a given number of visits within a specified
time span within a single night, followed by a given number of additional nights (each with the same required number
of visits in the same time span) falling within a specified time window. The basic criteria is a pair of visits in each
night occurring within 60 minutes, repeated for 3 nights within a 15 day time window. However, we also evaluate
the effect of varying the discovery criteria to require triplets or quads of visits within a single night, and increase
the length of the search window from 15 to 30 days. 

\subsubsection{Completeness Estimation}

With each unique set of discovery criteria, we have a record of what objects would be ``discovered'' at each $H$ value.
With this we calculate the differential discovery completeness, the fraction of objects discovered at a given $H$ magnitude.
To turn this into a cumulative discovery completeness, we simply integrate over $H$, assuming a given $H$ distribution
for the population (recall that we use $dN/dH = 10^{\alpha\, H}$, with $\alpha$ = 0.33).

\subsection{The LSST Baseline Survey\label{sec:surveys}}

The current baseline observing strategy for LSST is represented by our reference run, minion\_1016. This simulated survey
contains observations balanced between several different observing proposals as follows:
\begin{enumerate}
\item The Wide, Fast, Deep (WFD) proposal (also known as the Universal proposal) is the primary LSST survey, expected to receive about 90\% of the observing time and to cover 18,000 deg$^2$ of sky. In the baseline observing strategy, this proposal is configured to obtain visits in pairs spaced about 30 minutes apart, and will typically return to each field about every 3-4 days, balancing the six $ugrizy$ filters. The footprint for the WFD proposal covers approximately $+5^\circ$ to $-60^\circ$ in declination, with a full range of RA values except for a region around the Galactic plane. This declination range corresponds to an airmass limit of about 1.3 when the fields are at an Hour Angle of $\pm$2 hours. In minion\_1016, the WFD proposal receives 85\% (2,083,758) of the total number of visits.
\item The North Ecliptic Spur (NES) proposal is an extension to the WFD to reach the northern limits of the Ecliptic plane ($+$10 degrees), and allows higher airmass observations. The visit timing is similar to the WFD, although the $u$ and $y$ filter are not requested in this region. In the baseline observing strategy, minion\_1016, each NES field requests about 40\% of the total number of WFD visits per field when considering $griz$ filters only (304 visits per field in $griz$ vs 795 visits per field in $griz$ in WFD), and receives 6\% (158,912) of the total number of visits.
\item The Deep Drilling Fields (DD) proposal includes a set of single pointings that are requested in extended sequences; currently these sequences are $grizy$ visits, with additional coverage in $u$ band. Each sequence requires about an hour of observing time, and is repeated every few days. In minion\_1016, there are 5 DD fields, 4 of which correspond to fields which have been officially selected
by the Project and announced to the community; these five fields receive 5\% of the total visits.
\item The Galactic Plane (GP) proposal covers the region with high stellar density around the Galactic plane not covered by the WFD. This proposal requests a small number of visits in each of the six $ugrizy$ filters, with no timing constraints. In minion\_1016, this proposal receives 2\% of the total visits.
\item The South Celestial Pole (SCP) proposal is an extension of the WFD footprint to cover the region south of $-60^\circ$ declination. Like the GP, this proposal requests a small number of visits in each of the six $ugrizy$ filters, with no timing constraints. In minion\_1016, this proposal receives 2\% of the total visits.
\end{enumerate}

\begin{figure}[t!]
\centering
\includegraphics[width=0.85\textwidth]{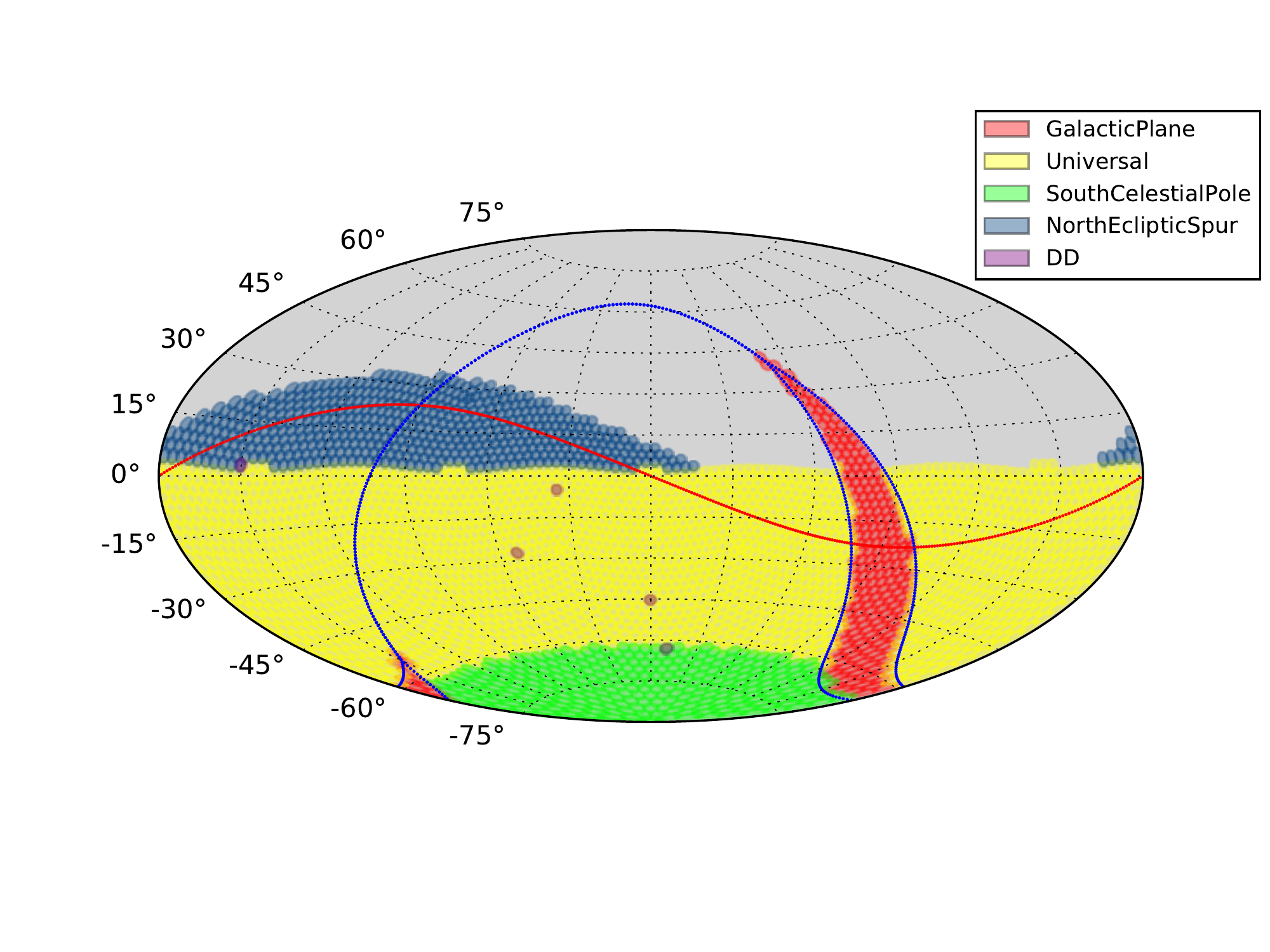}
\vskip -1.0in
\caption{The footprints of the various proposals included in the baseline observing strategy, represented by reference run minion\_1016. In the reference run, the WFD proposal receives 85\% of the total visits, the NES receives about 6\% of the
total visits, the DD receives about 5\% of the total visits, and the GP and SCP each receive about 2\% of the total visits. 
In the baseline run, every visit is 30s long, consisting of two back-to-back 15s exposures.
\label{fig:minion_footprints}}
\end{figure}

This is summarized in Table~\ref{tab:surveysummary}. The footprint of these various proposals in the baseline minion\_1016 reference run is shown in Figure~\ref{fig:minion_footprints}. In each proposal, the individual visits are 30 seconds long, consisting of two back-to-back coadded 15 second exposures.

\begin{deluxetable}{lccccc}
\tablecaption{Summary of the proposals in the baseline minion\_1016 simulated survey. \label{tab:surveysummary}}
\tablehead{Proposal & Filters & Pairs? & Number of fields & Number of visits/field & \% total visits}
\startdata
Wide Fast Deep & $ugrizy$ & Yes & 2293 & 908 & 85 \\
North Ecliptic Spur & $griz$ & Yes & 521 & 344 & 6 \\
Deep Drilling & $ugrizy$ & Yes & 5 & ~23,000 & 5 \\
Galactic Plane & $ugrizy$ & No & 230 & 180 & 2 \\
South Celestial Pole & $ugrizy$ & No & 293 & 180 & 2 \\
\enddata
\end{deluxetable}

\subsection{Baseline Survey Discovery Completeness}

\begin{figure}[t!]
\centering
\includegraphics[width=0.99\textwidth]{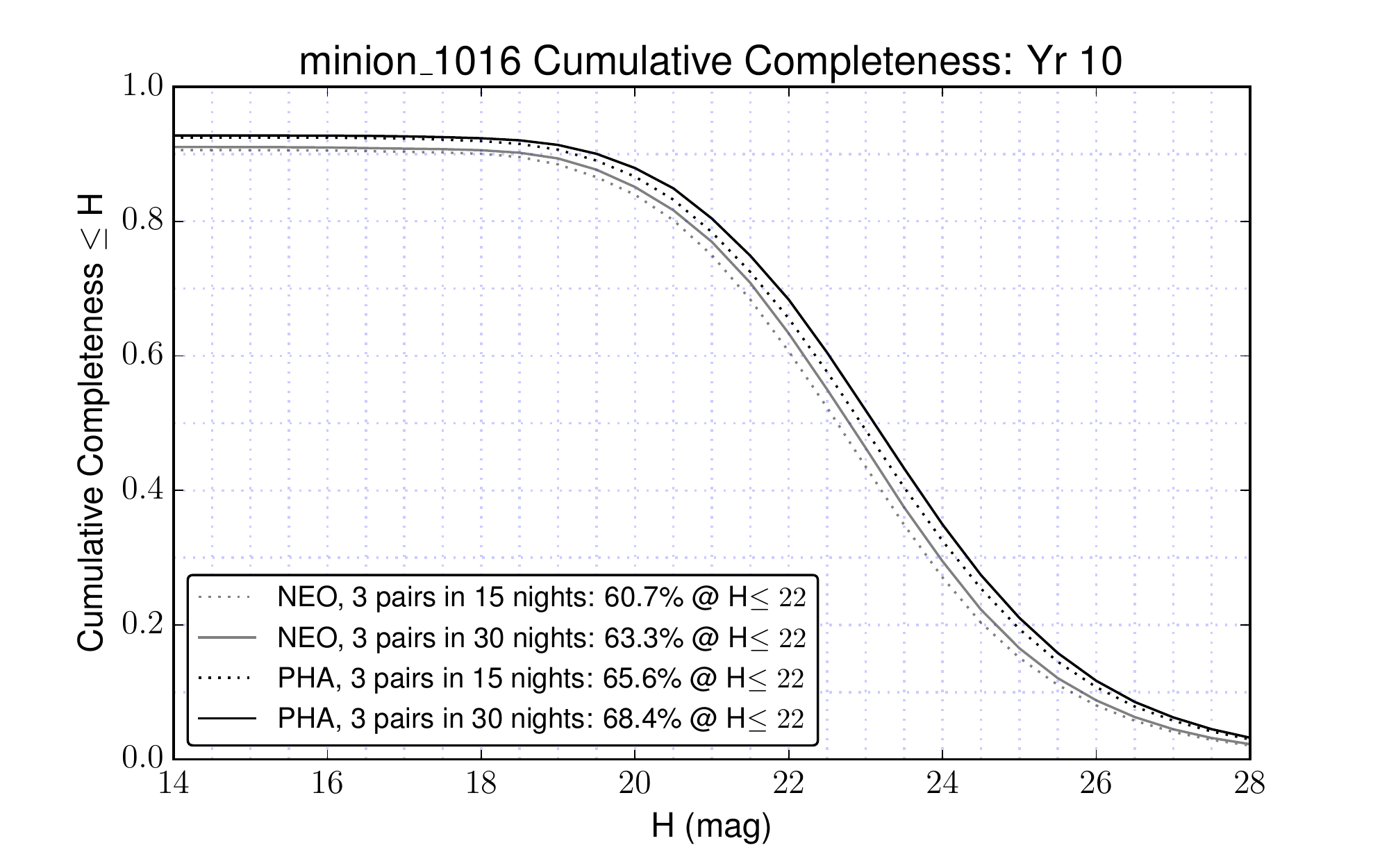}
\vskip -0.2in
\caption{The cumulative completeness for PHAs and NEOs, as a function of absolute magnitude $H$, for the baseline
cadence minion\_1016. The completeness is below 100\% at the bright end (large size limit) because some objects have
synodic periods longer than the survey duration (and thus effectively ``hide'' behind the Sun), and some are visible but
 do not receive the required number of observations due to telescope scheduling. \label{fig:minionC1}}
\end{figure}

Using the baseline survey run, minion\_1016, with the baseline discovery
criteria (pairs of visits occurring within 60 minutes and
repeated for 3 nights within a 15 day time window), we find a cumulative
completeness of
\begin{equation}
C_{\rm PHA, baseline}(H\le22) = 65.6\%
\end{equation}
at $H\le22$ for our PHA input population. This is LSST's baseline expected PHA completeness, derived using the reference cadence and the design MOPS and data management requirements. The cumulative completeness as a function of $H$ magnitude is shown in Figure~\ref{fig:minionC1}). 

When an NEO population is used instead of a PHA input population, the cumulative completeness is:
\begin{equation}
C_{\rm NEO, baseline}(H\le22) = 60.7\%
\end{equation}
or about 5\% lower than for the PHAs
(see also Figure~\ref{fig:minionC1}). This is primarily due to differences in the orbital distribution differences, as illustrated in Figure~\ref{fig:PHA_orbits}. The definition of PHAs includes a Minimum Orbit Intersection Distance (MOID) with Earth of 0.05~AU, requiring PHAs to more closely approach Earth than NEOs (which are defined as simply having $q<1.3$~AU), and thus the PHAs achieve brighter peak V magnitudes than the NEOs. To quantify this effect, we calculated the apparent $V$ magnitude for both the NEO and PHA input populations every night for ten years, while accounting for trailing losses and assuming a constant $H=22$ magnitude for every member of the population. The resulting mean value of the
brightest 10-year $V$ magnitudes are about half a magnitude brighter for PHAs than for NEOs.

Note that all the completeness results presented in this section assume that no objects are known prior to LSST survey, and thus are lower than estimates including other surveys. We'll return to the impact of adding discoveries by the  rest of the NEO discovery system in \S\ref{sec:known}. 

\begin{deluxetable}{lccccccc}
\tablecaption{The cumulative completeness for PHAs with $H\le22$ for various
survey strategies (rows) and MOPS discovery criteria (columns). The numbers presented here are for LSST alone, and do not include discoveries made by other surveys. In addition to changing the overall duration of the survey (12 years instead of 10), the
completeness is shown for different track linking windows ($N_w=15$ or $30$ days),
enhanced detection algorithms to reduce trailing losses (``Trail Det''), and
lowering the individual detection threshold from SNR$=5$ to SNR$=4$. These columns therefore map to software or computational capability enhancements, while the rows map to different survey cadences. See the text for detailed description of individual survey strategies. \label{tab:completeness}}
\tablehead{
                   & \multicolumn{3}{c}{10 year survey}  &  \multicolumn{4}{c}{12 year survey}  \\
    \cmidrule(r){2-4} \cmidrule(r){5-8}
Simulation  & $N_w$=15 & $N_w$=30 & $N_w$=30 & $N_w$=15 & $N_w$=30 & $N_w$=30 & $N_w$=30 \\
                   &                   &                    & Trail Det     &                   &                   &  Trail Det    &  SNR=4 
}
\startdata
LSST baseline & 65.6 & 68.4 & 69.1 &  --- & --- & --- & --- \\
Extra ecliptic visits & 66.1 & 69.8 & 70.5 & 70.5 & 73.9 & 74.8 & 77.1 \\
Longer ecliptic visits & 63.2 & 67.5 & 70.5 & 67.3 & 71.7 & 74.5 & 75.7 \\
NEO-focused cadence & 66.5 & 70.3 & 72.3 & 70.2 & 73.8 & 75.8 & 77.2 \\
\enddata
\end{deluxetable}

\subsection{Enhancing the Discovery Yields}

Given the baseline presented above, we can examine the effects of changing both the survey design (reallocating telescope resources) and improving the software and/or devoting more computing resources to the linking problem.


\subsubsection{Enhancing the Discovery Yields: Software}

The baseline analysis assumed that only objects linked in 15 day ``windows'' will be discovered, consistent with the design requirements and measured performance of LSST's implementation of MOPS. It is instructive to examine how much better the LSST would perform if further investment is made in enhancing the LSST software system (primarily MOPS, but image differencing as well).

Continuing to use the baseline minion\_1016 simulation, we explore the impact of possible software improvements on discovery rates. The results are detailed in Table~\ref{tab:completeness}. We find:

\begin{itemize}
\item {\bf 30-day window}: Extending the MOPS window for linking pairs of detections from the nominal 15 day window to a 30 day window increases completeness by about 3\%. This change also comes with an increase in computational requirements by about an order of magnitude (see Appendix \ref{sec:appMOPS}).

\item {\bf Detecting trailed sources}: Enhancing source detection algorithms to detect trailed $SNR=5$ objects increased completeness by about 0.5\% relative to the baseline, with little sensitivity to linking window size. This is because for most PHAs (and NEOs) LSST gets multiple discovery opportunities over the 10yr survey period.

\item {\bf Detecting at SNR=4}: Using sources detected down to SNR=4 instead of SNR=5 increases completeness by about 3.5\% relative to baseline. However, the change also leads to an estimated increase in the compute requirements by about two orders of magnitude (see \S\ref{sec:kaiser}).

\end{itemize}

Widening the MOPS tracklet linking window from 15 to 30 days achieves a meaningful gain in completeness relative to the baseline, with a reasonable marginal computational cost (discussed in \S\ref{sec:discussion}). The increased window allows more opportunities to capture a set of observations which meet the MOPS discovery criteria.

We also note that the current OpSim behavior does not prioritize capturing large chunks of contiguous sky, often leaving gaps in coverage from night to night. Therefore, with the large LSST field of view, after 30 days the areal coverage will be much more evenly distributed than after 15 days enhancing the 30-day-window yields.
Changes to the scheduling algorithm to favor covering contiguous blocks of sky\footnote{A similar modification of
the baseline cadence, the so-called ``rolling cadence'', is also favored by the supernovae science programs. A release of a series of simulated surveys implementing this idea is anticipated for late 2017.} are likely to further improve the completeness.

\begin{figure}[t!]
\centering
\includegraphics[width=0.99\textwidth]{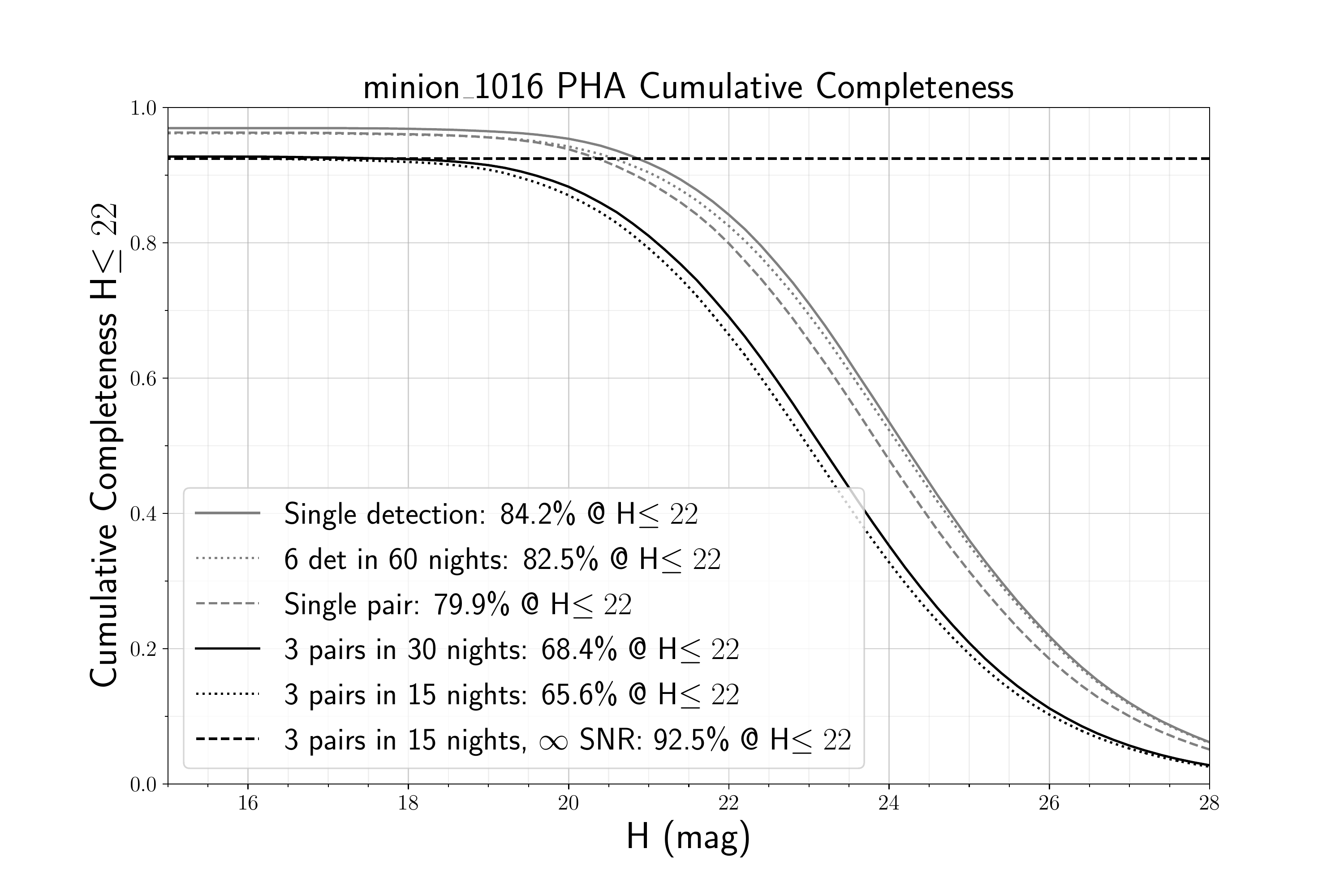}
\vskip -0.2in
\caption{The cumulative completeness for PHAs, as a function of absolute magnitude $H$, for the baseline
cadence minion\_1016, considering a variety of detection requirements. The completeness is below 100\% at the bright end (large size limit) because some objects have
synodic periods longer than the survey duration (and thus effectively ``hide'' behind the Sun), and also because some objects do not receive the required number of observations within the `window'. It is not due to the limited sensitivity of the LSST system, as can be seen by the ``3 pairs in 15 nights, $\infty$ SNR'' line, which shows the completeness expected when the discovery requirement is 3 nights with pairs of observations within a 15 night window but assuming an infinitely sensitive survey. The potential gains with better optimized scheduling (observing larger contiguous chunks of sky, for example), can be seen in the difference between the single pair of detections or 6 separate detections within 60 nights, vs. 3 pairs of detections in 15 or 30 nights (indicating potential gains of over 10\% in completeness for $H\le22$). \label{fig:more_completeness}}
\end{figure}

The upper limit to yield improvements due to software enhancements is likely $\gtrsim 10$\%; this is illustrated in Figure~\ref{fig:more_completeness}, where requiring only a single night of pairs or requiring 6 observations in any sequence over 60 nights increases completeness over 10\%.

\subsubsection{Enhancing the Discovery Yields: Survey Strategy}

\begin{figure}[t!]
	\centering
	\includegraphics[width=0.49\textwidth]{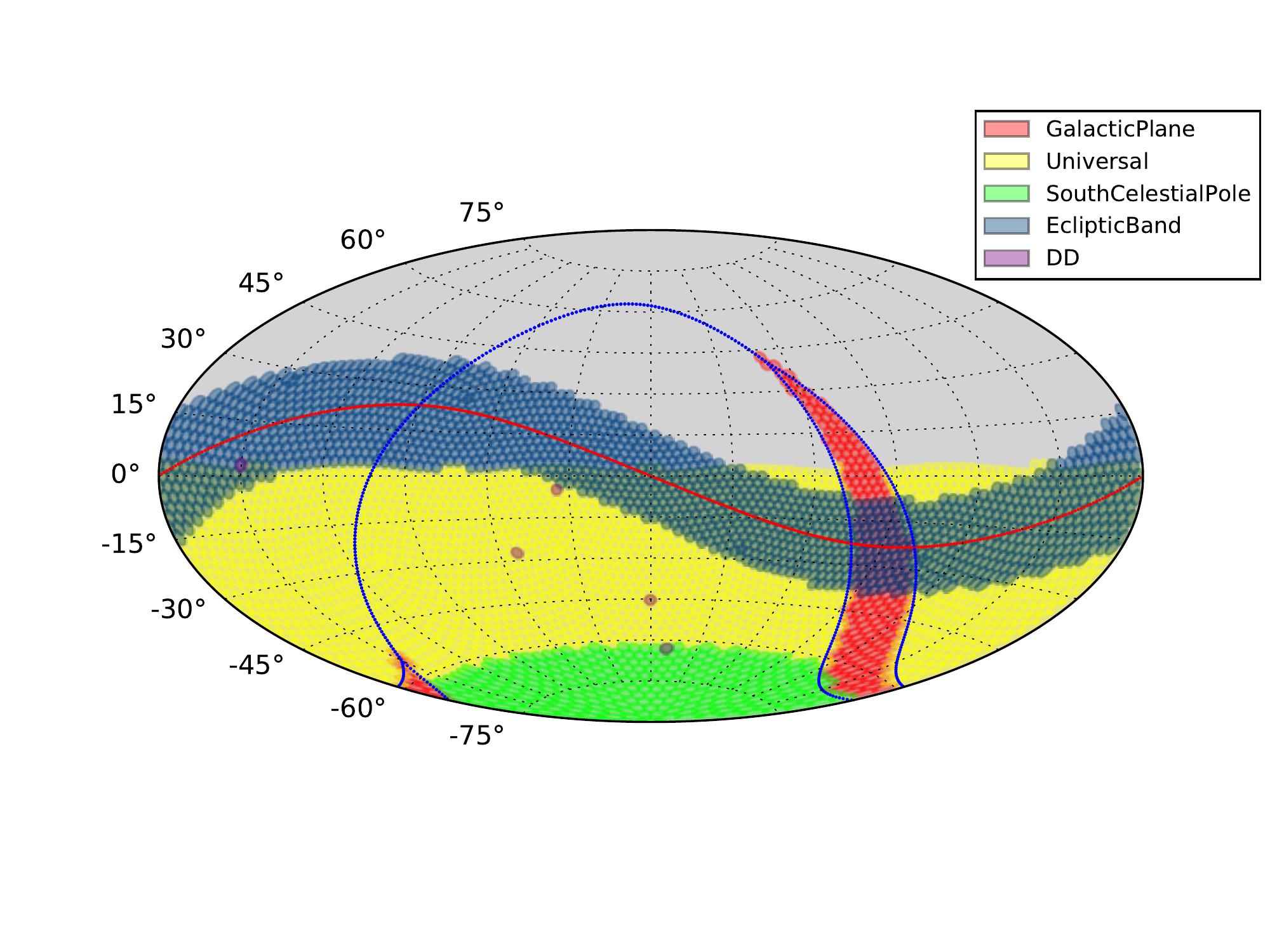}
	\includegraphics[width=0.49\textwidth]{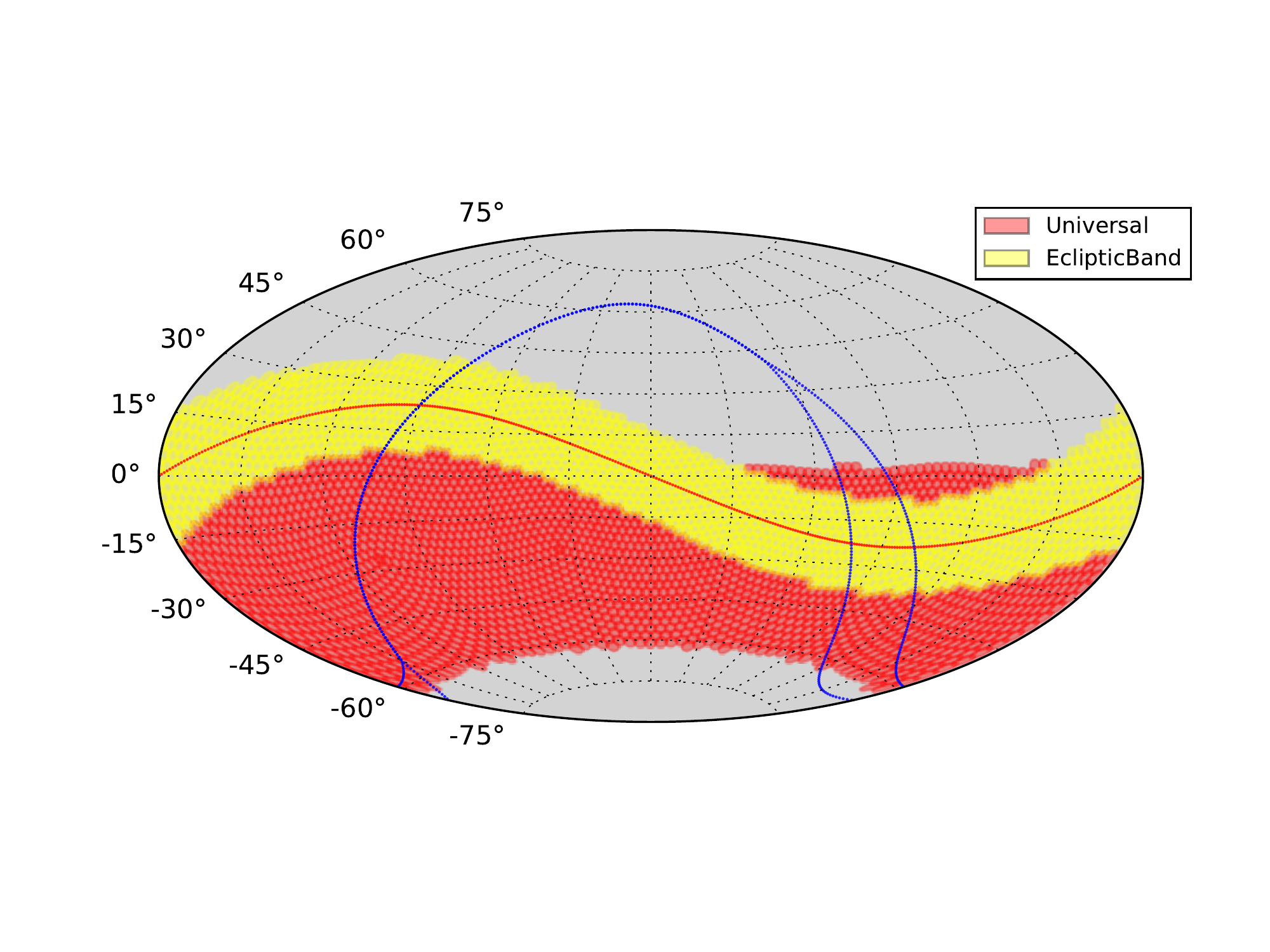}
	\vskip -0.5in
	\caption{The footprints of the proposals, including the Ecliptic Band proposal, used in the NEO-optimized simulated surveys astro\_lsst\_01\_1015 (the ``longer ecliptic visits'' survey, left) and astro\_lsst\_01\_1017 (the ``NEO-focused'' survey, right). The astro\_lsst\_01\_1017 survey only includes two proposals.
		\label{fig:neo_footprints}}
\end{figure}

We next look at modifying the survey strategy to increase LSST's NEO yields. We create a series of additional OpSim simulated surveys with parameters intended to improve the efficiency of discovering PHAs and increase the cumulative PHA completeness. These span the range from minor modifications to extreme changes that would jeopardize other LSST science goals. We consider the latter in order to assess what would be ultimate performance of an LSST-like system fully dedicated to NEO surveying.

The potential improvement in PHA discovery rates for modified survey cadences,
detailed in the rows of Table~\ref{tab:completeness}, is as follows:

\begin{itemize}
\item \textbf{Extra ecliptic visits:} By adding extra visits to the ecliptic spur (spending 24\% of survey time observing the NES proposal field, relative to 6\% in the baseline cadence), the increase in completeness over the LSST baseline is only about 0.5-1\%. This improvement comes at a cost to other science cases, as the main survey footprint (the WFD proposal) only receives 1,715,354 visits (82\%) of the number of visits in the reference run; the outcome of many science programs is roughly proportional to the number of visits.

\item \textbf{A 12-year survey:} By extending the baseline survey strategy by additional two years, we find an increase in completeness of 4\% over its completeness at the 10 year mark.

\item \textbf{Extra ecliptic visits, over 12 years:} We next look at the completeness of the ``extra ecliptic visits'' strategy, if extended to 12 years. We find it's better by $\sim 5$\% relative to the baseline.

\item \textbf{Longer visits in the ecliptic:} This strategy is visualized in the left panel of Figure~\ref{fig:neo_footprints}. The simulation requests longer, 60 second, visits in a band of $\pm 15^\circ$ around the ecliptic plane (the ``Ecliptic Band'' -- EB -- proposal).

The strategy reaches fainter limiting magnitudes, but the improvement due to longer exposures is counteracted by the fact that trailing losses are also increased. Similarly, the increase in the exposure time near the Ecliptic trades off against the revisit frequency reducing the number of opportunities for sets of observations that MOPS can successfully link. 

Because of that, this survey strategy show a slight {\em decrease} in completeness (percent level), relative to the baseline processed with the same software. Only if we assume the object detection pipelines can be modified to optimally detect trailed objects, a slightly higher PHA completeness level (~0.5\%) is achieved.

\item \textbf{NEO-focused survey:} This survey is visualized in the right panel of Figure~\ref{fig:neo_footprints}. It uses a limited filter set, discards other proposals, and uses longer exposures along the ecliptic. This strategy shows a slight increase in completeness ($\sim1-3$\%) relative to the baseline.
However, other LSST science programs would be jeopardized with this observing strategy because observations in the $uzy$ filters,
as well as special program observations would not be obtained.
\end{itemize}

To summarize, when altering the survey strategy the largest individual gain ($\sim$4\%) comes from simply extending the survey duration from 10 to 12 years. Other tested variants result in smaller marginal improvements, yet negatively affect other LSST science cases (most severely so in the case of the aggressive NEO-focused strategy).

\subsection{LSST in the Context of the Broader NEO Discovery System\label{sec:known}}

The completeness estimates presented above assumed that no objects are known prior to LSST survey or 
discovered by other resources during the LSST lifetime. When those contributions are taken into account, the completeness achieved by the entire system is significantly higher than what LSST alone can deliver.


We estimate the discoveries contributed by these previous, current, and future resources 
using a model similar to \citet{VeresChesley2017neo}: 
after generating ephemerides for the same populations of NEOs and PHAs as above, daily from 2000--2034 
(the end of the extended LSST surveys), we clone the objects over the same range of $H$ as 
used in the discovery metrics above. Each clone is considered potentially discovered if, while it is at
a solar elongation greater than 100 degrees, it becomes brighter a given apparent magnitude threshhold: 
this threshhold is $V=20$ from 2000-2005 (corresponding to the LINEAR era), $V=21$ from 2005-2015 
(representing Catalina and PanSTARRS1), $V=22$ from 2015 to the start of LSST in 2022, and then
$V=22.2$ after 2022.  In each of these periods, it is possible that even if the objects are bright enough, 
they would not be discovered simply due to sky coverage constraints of the telescopes; to 
this we add an efficiency factor that accounts for how likely any given detection on a given night is to be achieved. 
This factor is held constant until 2022, and then doubled (together with the increase in magnitude 
limit) to account for future improvements in NEO survey resources. We tune the efficiency factor to match
the historical rate of discoveries reported on the JPL NEO discovery page\footnote{See https://cneos.jpl.nasa.gov/stats/totals.html}.

With this model, at the start of LSST, we estimate the known NEO population with $H<22$ to be 43\% complete
and the PHA population with $H<22$ to be 54\% complete. Our NEO estimate is in good agreement with those
from \citet{VeresChesley2017neo} and \citet{GMS2016} who predict 42\% and 43\% completeness for NEOs, respectively.
In 2032, at the nominal end of LSST, our model predicts 72\% completeness for PHAs and 59\% completeness for NEOs, all without LSST. For NEOs, \citet{VeresChesley2017neo}
predicts a slightly higher value of 61\% while \citet{GMS2016} predicts a lower value of about 52\%. 
Given the uncertainty in the assumptions being made about the evolution of other surveys over a 17 year time period, this variance is understandable.

Combining these discoveries with those in the baseline cadence ({\it minion\_1016} with 15 day windows) 
gives us the expectation for our knowledge of the the post-LSST completeness of PHAs and NEOs. We find:
\begin{equation}
C_{\rm PHA, baseline + system}(H\le22) = 81\%
\end{equation}
and
\begin{equation}
C_{\rm NEO, baseline + system}(H\le22) = 73\%
\end{equation}.
In other words, the Earth-approaching asteroids discovered by the LSST will add $\sim 10$ percentage points to what the system would have discovered (given our model assumptions).

\begin{figure}[t!]
\centering
\includegraphics[width=0.99\textwidth]{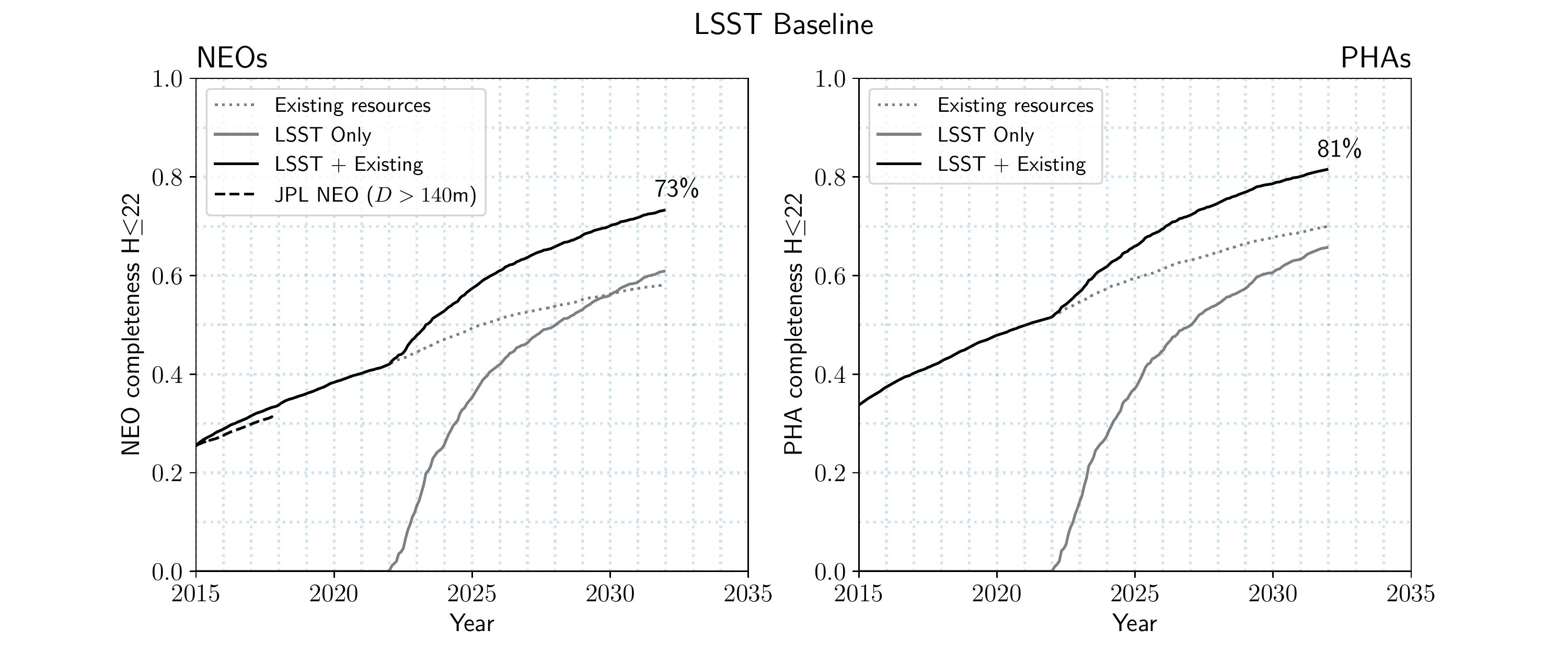} \\
\includegraphics[width=0.99\textwidth]{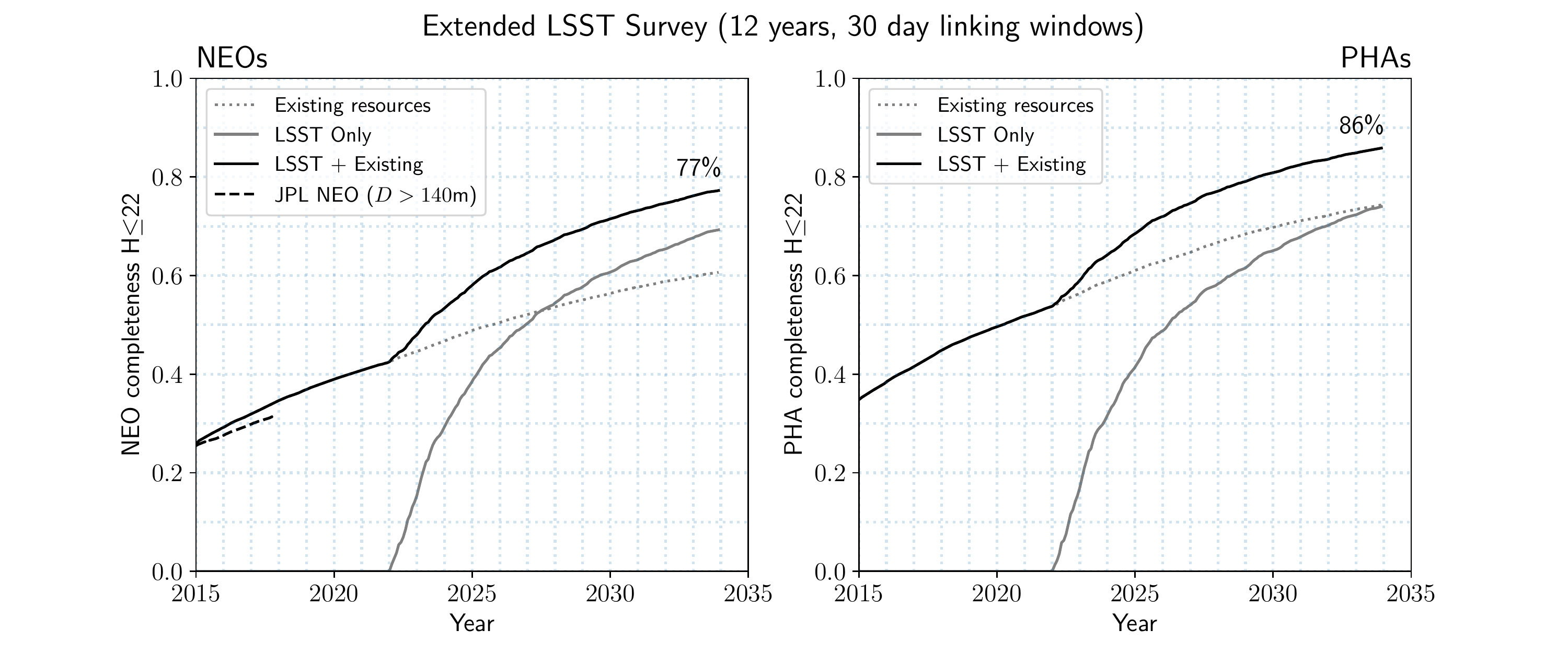}
\vskip -0.2in
\caption{The cumulative completeness for NEOs (left) and PHAs (right) with $H\le22$, as a function of
time, when known objects (gray solid lines) are taken (black solid lines) and not taken (dotted lines)
into  account (see also Table~\ref{tab:completeness2}). Discovery with existing and future resources 
is accounted for with a model that attempts to match expected limited magnitudes in each time period
and an efficiency factor that is tuned to match the rate of discoveries as reported by JPL (dashed line; see text for details). 
The top panel shows discoveries assuming the baseline LSST survey (10 year duration, with a 15 night tracklet linking window).
The bottom panel shows discoveries that could be achieved with the 12 year ``extra ecliptic visits'' LSST survey and a 30 night tracklet linking window.
\label{fig:knownObj}}
\end{figure}

Combining these discoveries with those from the ``extra ecliptic visits'' strategy under a 30-day MOPS linking window, we find that including the LSST into the NEO discovery system boosts overall PHA completeness to 84\% over a 10-yr period, and 86\% if the survey is extended to 12 years. These results are summarized in Figure~\ref{fig:knownObj} and Table~\ref{tab:completeness2}.

\begin{deluxetable}{lcccccc}
\tablecaption{The cumulative completeness (in \%) for NEOs and PHAs with $H\le22$ for
the 10 year LSST baseline survey strategy, with a linking window ($N_w$) of 15 days,
compared with an extended 12-year survey with additional visits along the
Ecliptic using 30 day linking windows. The estimate of completeness achieved by other surveys
without LSST is given in ``No LSST'', the estimate of LSST discoveries alone is given in ``Only LSST'', 
and the combination of LSST plus the larger discovery effort is given in ``LSST + others''. 
\label{tab:completeness2}}
\tablehead{
& \multicolumn{3}{c}{10 year baseline, $N_w$=15}  &  \multicolumn{3}{c}{12 year ``extra ecliptic visits'', $N_w$=30}  \\
\cmidrule(r){2-4} \cmidrule(r){5-7}
Population  & No LSST &  Only LSST & LSST + others &  No LSST  & Only LSST & LSST + others
}
\startdata
    NEO & 59 & 61 & 73 & 61 & 69 & 77 \\
    PHA & 72 & 66 & 81 & 74 & 74 & 86 \\
\enddata
\end{deluxetable}

\subsection{Systematic Effects due to Varying Modeling Assumptions \label{sec:syseff}}

As indicated by the above discussion, a number of systematic effects must be taken into account when
comparing different simulations of the same survey, as well as simulations of different surveys and observing
systems. Below we list the leading systematic effects in simulated completeness estimates:
\begin{enumerate}
\item Population choice - an NEO vs. PHA difference. We find the completeness is generally about $\sim$5\% higher for PHAs than for NEOs.
\item Orbital parameter distribution for the simulated asteroid population (e.g. the \citealt{Bottke2002} model
             vs. the Granvik, in prep., model). Varying populations contribute completeness uncertainty of about a few percent, based
             on simulations by \citet{VeresChesley2017neo}.
\item Different sample definitions: $H<22$ vs. $D>140$m; as shown by  \citet{2016AJ....152...79W} and \citealt{GMS2016},
          completeness increases by $\sim$5 percentage points (p.p.) when an $H$-based criterion is used instead of $D>140$m.
\item Variations of the ``discovery window'' (e.g., 3 nights with pairs of visits over $N_w$ nights: increasing
        $N_w$ from 15 to 30 increases completeness by about 3-5 p.p., while decreasing $N_w$ from 15 to 12 decreases 
        completeness by about 1-2 p.p. depending on the details of the survey cadence).
\item Uncertainties when predicting effective image depth (system throughput, background noise due to sky brightness, 
          variation of the detection efficiency with the signal-to-noise ratio, treatment of trailing losses). As a rule of thumb, for a survey that has a 
          completeness above 60\%, each additional 0.1 magnitude of depth for a given survey cadence increases the 
          completeness by another 1 p.p.
\item Uncertainties when predicting asteroid's apparent flux (albedo distribution, phase effects, photometric variability
          due to non-spherical shapes, color distributions); assuming an uncertainty of 0.2 mag in the effective
          limiting magnitude, the corresponding  systematic uncertainty in completeness is about 2 p.p.
\item Variations of the nominal detection threshold. If the detection threshold is changed from the
          signal-to-noise ratio of 5 or greater to 4 or greater, the completeness is boosted by $\sim$3~p.p.;
          the difference between the optimal detection using trailed profile and point-spread-function
          detection, which is negligible for LSST baseline exposure time of 30 seconds, would be worth $\sim$1.5~p.p.
          in completeness for visits with a doubled exposure time.
\item Sensitivity to details in sky coverage and cadence (e.g. nightly pairs of visits vs. quads of visits).
          Requiring quads instead of pairs of visits decreases completeness by 30\% using the baseline cadence;
          about half of that loss can be recovered using cadence simulations that request four visits per night.
\item The slope of the asteroid size distribution. Current measurement uncertainty of this parameter
          corresponds to a systematic uncertainty in completeness of about 2\%.
\item The impact of known objects. As discussed in \S\ref{sec:known}, we estimate that 54\% of PHAs with $H<22$ would be discovered
          by current survey assets by the start of LSST survey in 2022 (currently $\sim$36\%), and they would
          boost the final PHA completeness after a 10-year LSST baseline survey by 14~p.p. But different models and assumptions on the future development of the NEO system make this number uncertain to at least $\sim 3-5$~p.p.
\end{enumerate}

Given all these, it is unlikely that a meaningful quantitative comparison of different studies can be expected beyond a level of a few percent. The same is true of the accuracy of an estimate vs. actual realized performance. The exact magnitude of that uncertainty is difficult to estimate as the various effects don't add up coherently, aren't equally likely, with some being poorly constrained. Effects 1, 3, 4, 7, and 8 are a matter of definition/input. Adding all other effects in quadrature yields a notional ``error bar'' of $\sim 5$ percentage points, which we adopt here.

\subsection{Comparison to Other Works  \label{sec:other}}

Two recent works, \citet*[][hereafter GMS]{GMS2016} and \citet*[][hereafter VC]{VeresChesley2017neo}, 
have also evaluated the NEO completeness that could be achieved with a baseline LSST survey.
GMS reported a cumulative completeness for NEOs of 63\%, 62\% for PHAs,
while VC reported a cumulative completeness of 61.6\%  for NEOs and 64.9\% for PHAs, 
for a baseline survey with $N_w$ = 15 days. These can be compared with our estimates of 61\% for NEO completeness
and 66\% for PHA completeness.

All these results are consistent, given the understood systematic differences discussed in \S\ref{sec:syseff}. We identify three main reasons that contribute to the (small) differences:
\begin{enumerate}
\item The definition of the completeness limit: $H\le22$ or $D>140$m. By using an albedo distribution combined 
with a size distribution, GMS directly model the population larger than 140m rather than those with $H\le22$. 
Using $H\le22$ can lead to a increase in completeness on the order of 5\% \citep{2016AJ....152...79W, GMS2016}.
\item The realization of the LSST baseline survey, including the sky brightness model, system sensitivity model, and
delivered seeing distribution.  GMS used an older realization of the LSST baseline survey, {\it enigma\_1189}, which
had limiting magnitudes that were on average fainter by a few tenths of a magnitude than they are in 
{\it minion\_1016}, a more recent baseline survey with updated values for the system throughput and seeing. 
The decrease in limiting magnitudes in {\it minion\_1016} leads to a decrease in completeness of about 2\% compared
with GMS results.
\item Different realizations of the input populations. The major difference here is between NEO and PHA populations,
which leads to the difference in completeness in our results and the results from \citet{VeresChesley2017neo}.
However, small differences in the source of the input population can also lead to differences on the level of a few percent \citep{VeresChesley2017neo}. 
\end{enumerate}

Therefore, after accounting for different choices of simulation parameters 
we conclude the GMS and VC results for baseline NEO discovery efficiency are in excellent mutual agreement (within 1-2 \%), as well as in agreement with the analysis presented in this paper.

\section{Discussion and Conclusions\label{sec:discussion}}

We have tested and quantified the ability of LSST to contribute to the census of NEO and PHA populations, as well as options for further improvement of LSST's NEO and PHA yields relative to the baseline cadence.

We expect that the LSST strategy for discovering moving Solar System objects will be successful because the following three conditions are likely to be met:
\begin{enumerate}
	\item The LSST system hardware and image differencing software performance will result in false detection
	rates not significantly exceeding $\rho_{FP} =  450$ deg$^{-2}$, conservatively estimated here using real DECam data
	processed using prototype LSST software.
	\item Given an anticipated 1000-core machine, LSST MOPS will be able to easily process as many as
	10$^8$ tracklets per search window, and daily computations to produce up to about 10$^7$
	candidate tracks will be completed in about an hour. We demonstrate this by running an LSST MOPS prototype on a representative simulated dataset. It has been independently confirmed by \cite{VeresChesley2017mops}, using Pan-STARRS1 MOPS.
	\item With the IOD computational budget of 0.1 sec per track -- comfortably above the value of $<26$ms we measure here -- the final track filtering step can
	be easily accomplished in about an hour.
\end{enumerate}

Our determination of the expected false detection rate is based on processing data acquired by DECam using prototype LSST pipelines. With a conservative extrapolation of these data to expected LSST depth, we find expected rates of false detections of $450$~deg$^{-2}$. This estimate includes no provision for real-bogus type classifiers, which have been successfully applied by existing surveys to reduce the false detection rates by an order of magnitude or more \citep[e.g.][]{goldstein15}. It is therefore best to think of it as a {\it conservative upper limit}, possibly overestimating the true rate by a factor of few.

Assuming this rate, numerical tests with LSST MOPS prototypes and representative IOD routines demonstrate that the planned compute system will be (more than) adequate to
process LSST data. Even if the realized false detection rate is twice as high as
the (already conservative) estimate reported here, it can still be handled without a change of baseline cadence, linking criteria, or
increase in computing resources. Quantitatively, the false detection rates of up to about
1000 deg$^{-2}$ can be readily handled with an approximately 1000-core cluster dedicated to moving object processing\footnote{Based on LSST's technology estimates and the data center sizing model, this equates to just $36$ CPUs, or $18$ compute nodes, in late 2022 at the beginning of the survey.}.


Having established that the LSST will successfully identify moving objects, we've examined the expected discovery yields for the PHA and NEO populations. Our analysis shows that LSST alone using the baseline survey strategy would discover about 66\% of PHAs with $H<22$ (and 61\% of NEOs). When objects expected to be discovered by prior and contemporaneous surveys are added, we find 81\% of all PHAs (73\% of NEOs) will be known by 2032. Expressed differently, adding the LSST to the global NEO discovery system will reduce the population of remaining unknown PHAs by 32\% (34\% for NEOs).

We compared the results presented here to two analogous recent studies by \citet[]{GMS2016} and \citet{VeresChesley2017neo}, and find them consistent within the modeling uncertainties (we note that due to population unknowns and other systematics, all these estimates are likely accurate and comparable to about $\pm 5$ percentage points). The original PHA completeness forecasts published in \cite{IvezicNEO2007} are higher than reported here (75\% vs 66\%), primarily due to using different PHA populations and assuming a 30-day search window in that work.

We examined a number of possible enhancements to the LSST baseline survey that could lead to further improvements. We find that the ``extra ecliptic visits'' observing strategy ({\it astro\_lsst\_01\_1016}), executed over 10 years and with a 30-day MOPS linking window could boost the overall PHA completeness to 84\% (including objects discovered by contemporaneous surveys). The downside to this option is that the remaining LSST science cases would receive $\sim 20$\% less observing time, likely rendering it unacceptable to the broader LSST community. 

However, if the survey is extended to 12 years and utilizes this strategy, the main LSST proposal (``deep-wide-fast'') would receive the same number of visits as with the 10-year baseline cadence ({\it minion\_1016}). Also, the PHA completeness would rise to 86\% (77\% for NEOs) for a reduction of the remaining unknown PHA population by an additional 26\% relative to the baseline (15\% for the NEOs). This makes this option particularly attractive as, assuming resources were identified to extend the survey by two years, the fraction of discovered PHAs would approach $\sim 90$\% without seriously disadvantaging other LSST science cases.

\acknowledgements
This material is based upon work supported in part by the National Science
Foundation through Cooperative Agreement 1258333 managed by the Association of
Universities for Research in Astronomy (AURA), and the Department of Energy
under Contract No. DE-AC02-76SF00515 with the SLAC National Accelerator
Laboratory. Additional LSST funding comes from private donations, grants to
universities, and in-kind support from LSSTC Institutional Members. MJ and CTS wish to acknowledge the support of the Washington Research Foundation Data Science Term Chair fund, and the University of Washington Provost’s Initiative in Data-Intensive Discovery.

\appendix
\section{The Impact of False Detections on MOPS Performance \label{sec:appMOPS}}

We seek to develop an analytic understanding for the behavior of the MOPS results.
In particular, we want to be able to predict the numbers of tracklets and
candidate tracks for a given input number of true and false detections. In addition, we seek
to understand how these numbers scale with the search window width,
velocity cutoff when forming tracklets, the temporal separation of two
detections in a tracklet, and the density of false detections. For example, available
MOPS experiments indicate that the number of tracklets increases with
the square of the false detection density, but other scalings are unclear,
especially the behavior of false candidate tracks.

We first derive the simpler false tracklet rates, and then use these results to
discuss false candidate track rates.

\subsection{Expected False Tracklet Rates \label{sec:tracklets} }

Given a detection in the first difference image, another difference image, obtained at a different epoch,
is searched for a matching detection to form a tracklet,  {\it e.g.} \citet{denneau13, kubica07}. For orientation,
the sky density of asteroids down to LSST $5\sigma$ faint flux limit ($r \sim 24.5$) is of the order
$\rho_{ast} \sim 100$ deg$^{-2}$. The predicted highest asteroid sky density for $r<24.5$,
on the Ecliptic, is up to about five times larger (with an uncertainty of about a factor of 2,
depending on model assumptions), and the density decreases rapidly with the ecliptic latitude.
A typical LSST observing night includes about 1000 visits, with two visits per night over
the active sky area. The nominal LSST field-of-view area is $A_{FOV}=9.6$ deg$^2$, with a
fill factor of 0.9, giving an effective field-of-view area of $A_{FOV}^{eff}=8.64$ deg$^2$. Hence,
the number of detected asteroids per night is of the order 500,000 (with implied two detections
per asteroid), although it can be significantly lower when the Ecliptic is not well covered (and
it could be a few times higher if the majority of visits were obtained along the Ecliptic).

The number of false detections due to (Gaussian) background fluctuations is
about $\rho_{bkgd} = 60$ deg$^{-2}$, assuming typical LSST seeing (0.8 arcsec)
and SNR$>$5. For a given seeing and SNR threshold, the rate of false detections can never be
lower than this estimate. This false detection rate decreases with the square of the seeing, and
strongly depends on SNR: the rate increases/decreases by as much as a factor of about ten
when SNR threshold is changed to 4.5 and 5.5, respectively (see \S\ref{sec:imDiff}).

Analysis of DECam images reduced using prototype LSST software, described in \S\ref{sec:imDiff},
shows a higher rate of detections in difference images, and a fraction of those detections
cannot be readily associated with true moving objects. This analysis implies a conservative
upper limit for the false detection rate of about $\rho_{FP} =  400$ deg$^{-2}$. This value
is conservative because analyzed DECam fields are close to the Ecliptic, with a significant but
not well known contribution from real asteroids (due to very faint flux levels, $r \sim 24$),
and it also includes true astrophysical transients that are not associated with static objects
(stars and galaxies). It is quite possible that the false detection rate might be several
times lower, though we will proceed with the most conservative estimate above.

The sky density of detections in difference images, $\rho_{det}$, is given by
the sum of contributions from true asteroids and false detections, $\rho_{det} = \rho_{ast} + \rho_{FP}
= 500$ deg$^{-2}$. When searching for a matching detection in another difference image, there are
two distinct types of behavior. Correct matches of detections of the same asteroid into tracklets follow the behavior
expected for correlated samples: as long as the object's angular displacement between the two epochs
is sufficiently larger than the seeing disk, while at the same time smaller than the search radius, the
number of matches (that is, the number of true tracklets produced per LSST pointing, assuming
two visits of the same area per night) is simply
\begin{equation}
                  N_{tracklet}^{true} = \rho_{ast}  \, A_{FOV}^{eff},
\end{equation}
With  $\rho_{ast} = 100$ deg$^{-2}$, $N_{tracklet}^{true} \sim 1,000$ per a pair of visits, and with
500 visit pairs per typical observing night, $N_{tracklet}^{true} \sim 500,000$ per night (same as
the number of detected asteroids in the active sky area, of course). Again,
this number can be much lower for fields far away from the Ecliptic, and a few times larger
for exceptionally good coverage of the Ecliptic. We emphasize that this number of true tracklets
does not directly depend on the search radius, nor the time elapsed between the two visits, as long
as they have their plausible values (about an arcminute, and a few tens of minutes, as discussed
further below).

\begin{figure}[t!]
\centering
\vskip -2.0in
\includegraphics[width=0.95\textwidth]{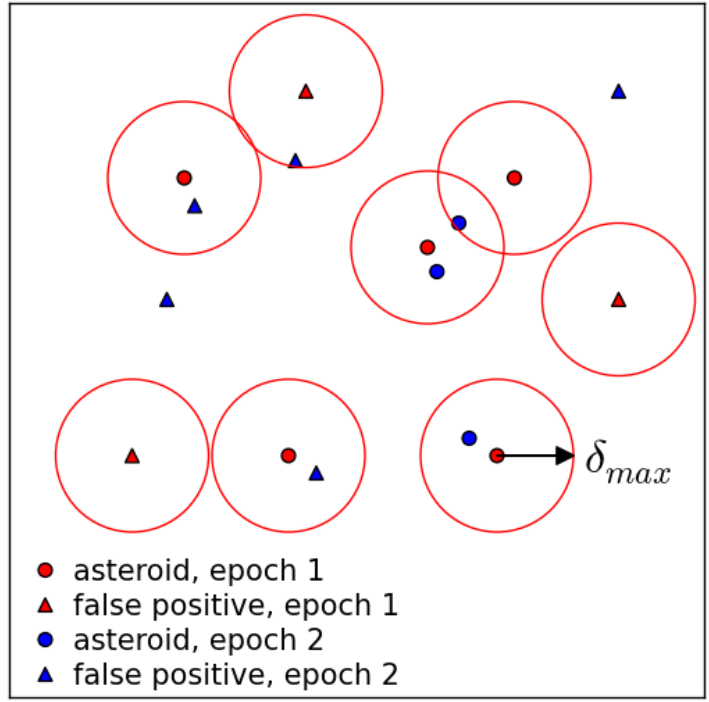}
\vskip -2.2in
\caption{An illustration of positional matching of detections to form tracklets.
Detections come in two flavors: asteroids (A, circles) and false detections (FP, triangles).
The figure shows the search for a matching detection in epoch 2 for each detection
in epoch 1, with a maximum search radius $\delta_{max}$. Note that there are six
possibilities: matches A-A, A-FP, FP-A, FP-FP, and orphaned A and FP.
\label{fig:TrackSlide2}}
\end{figure}

There are three other types of tracklets that follow the behavior for uncorrelated (random)
samples: associations of different asteroids, associations of asteroids and false detections,
and tracklets made of two false detections. Assuming the same $\rho_{det}$ in both
difference images, for each of $N_{det} = \rho_{det} \, A_{FOV}^{eff}$ detections in one image,
we search for a matching detection in another image (see Figure~\ref{fig:TrackSlide2}).
The search radius is given by
\begin{equation}
                     \delta_{max} = v_{max} \, \Delta t .
\end{equation}
Here $v_{max}$ is the  cutoff velocity and $\Delta t$
is the time elapsed between the two images. For LSST baseline cadence, $\Delta t$ is in
the range 20-60 minutes. The search area, $A_S = \pi \delta_{max}^2$, is then
\begin{equation}
\label{eq:AS}
      A_S = 0.0055 \left( v_{max}  \over {\rm deg \, day}^{-1} \right)^2 \, \left(\Delta t \over {\rm hour} \right)^2 {\rm deg}^2.
\end{equation}
To guide setting the cutoff velocity, simulations imply that 95\% of NEO detections have $v<1$ deg day$^{-1}$; with this threshold,
the completeness for main-belt asteroids is essentially 100\%. Objects moving faster than 1 deg day$^{-1}$ will
be easily resolved in LSST images and can be treated separately using specialized algorithms.
Adopting $v_{max} = 1$ deg day$^{-1}$,  and $\Delta t = 30$
minutes (which together  imply a search radius of $\delta_{max} = 1.3$ arcmin), gives a search area of
$A_S = 0.0014$ deg$^2$.

The expectation value for the number of matching detections within the search area $A_S$ (that is, the expected
number of tracklets per matching trial) is
\begin{equation}
                      p_{tracklet}^{false} =   \rho_{det}  \, A_S,
\end{equation}
and the total expected number of {\it false} tracklets for $N_{det}$ trials is thus
\begin{equation}
\label{eq:NttFalse}
           N_{tracklet}^{false} = N_{det} \, p_{tracklet}^{false} =  N_{visit} \, \rho_{det}^2 \, A_S \, A_{FOV}^{eff} = N_{visit} \, \rho^2_{FP}  \, A_S \, A_{FOV}^{eff} \,  \left(1 + 2 \eta + \eta^2\right),
\end{equation}
where $\eta = \rho_{ast}  / \rho_{FP} \sim 0.25$ (recall that $\rho_{det} = \rho_{ast} + \rho_{FP}$).
With $\rho_{ast} = 100$ deg$^{-2}$ and  $\rho_{FP} = 400$ deg$^{-2}$,
$N_{tracklet}^{false} \sim 3,000$ per pair of visits, and $N_{tracklet}^{false} \sim 1.5$ million per observing night with
$N_{visit}=500$ visit pairs. We note that the density of false tracklets ($\rho_{tracklet}^{false}=350$ deg$^{-2}$) is similar to
$\rho_{FP}$; this similarity is a consequence of choosing $\delta_{max}$ such that $\rho_{FP} A_S \sim 1$.

The first term in eq.~\ref{eq:NttFalse} is the largest and corresponds to tracklets made of two false
detections ($\sim1.0$ million), the second term corresponds to associations of asteroids and false detections,
and the third and the smallest term ($<0.1$ million) is due to incorrect associations of different asteroids.
For the chosen parameter values, the total number of tracklets is about 2 million per observing night, Given that
these choices are rather conservative, this estimate is essentially an upper limit; approximately,
{\it we expect of the order a million tracklets per observing night}.

To the first order ($\eta \approx 0$), the total number of tracklets per night is
\begin{equation}
    N_{tracklet} =  N_{tracklet}^{true} + N_{tracklet}^{false} =
       N_{visit} \, A_{FOV}^{eff} \, \left(\rho_{ast}  + \rho^2_{FP}  \, A_S \right).
\end{equation}
In addition to $N_{tracklet}^{false}$ scaling with the square of $\rho_{FP}$, as demonstrated using MOPS,
$N_{tracklet}^{false}$ scales with the square of
both $v_{max}$ and  $\Delta t$ (via the dependence on $A_S$). Therefore, if $\Delta t$ would be made
as small as 10 minutes by modifying the observing strategy, the resulting $N_{tracklet}^{false}$ would be about an
order of magnitude smaller (and $N_{tracklet}$ about three times smaller).  Hence, the shortening of $\Delta t$ is
a good mitigation strategy against high false detection rates in difference images\footnote{An
extreme example of this mitigation strategy would be to obtain two consecutive 30-second visits -- their
mid-exposure times would be separated by 34 seconds (additional 2 seconds due to shutter motion and another
2 seconds due to readout), which is sufficient to detect motion faster than about 0.1 deg day$^{-1}$.}.

\subsubsection{False tracklet velocity distribution \label{sec:falsev}}

False tracklets have randomly distributed velocities (motion vectors) with a cutoff given by $v_{max}$
(recall that $v_{max} = 1$ deg day$^{-1}$ was adopted above). The implied tracklet velocity is given by
\begin{equation}
                       v =  \delta / \Delta t,
\end{equation}
where $\delta$ is the angular separation of two detections. Since the number of tracklets
with separation $\delta$ increases linearly with $\delta$ (because the area of a circular
annulus is $2\pi r dr$), the false tracklet velocity distribution will increase linearly with
$v$ for $v<v_{max}$, and the vector orientation will be random. We show below that candidate
tracks can be efficiently pruned using this result.

\subsection{Expected False Track Rates \label{sec:tracks} }

In this section, we present an approximate estimate of the expected number of false candidate tracks.
Our goal is to derive the scaling of this number with the relevant input parameters, such as the true and false
tracklet rates per night ($N_{tracklet}^{true}=5\times10^5$ and $N_{tracklet}^{false}=1.5\times10^6$, respectively).
For a fiducial case, we assume that the search window is $N_w= 30$ days wide;  therefore, with $N_{tracklet} = 2\times10^6$
per night, there are $6\times10^7$ tracklets in the fiducial dataset. With about 4,300 deg$^2$ (500 pairs of visits)
of sky observed each night, the average density of (all) tracklets is $\rho_{tracklet} = 450$ deg$^{-2}$. Assuming
that on average the same field is revisited every $T_{revisit}=3$ days, the active area includes about 13,000 deg$^2$
of sky.

As discussed below in more detail, there are of the order 1000 different ways to chose a triplet of nights
from the search window. Given 10$^6$ tracklets per night, there are of the order
10$^{21}$ different combinations of tracklet triplets that could form a candidate track.
While this number of candidate tracks is obviously prohibitively large to test for consistency
with heliocentric Keplerian motion, it can be sufficiently reduced (to about the same number
as the number of true tracks along the Ecliptic) using pre-filtering steps based on tracklet motion
vectors, summarized below and following similar methods as described in \citet{denneau13, kubica07}. 

In the first step, the motion vector of a tracklet from the first night is linearly extrapolated
to the second night and tracklets from the second night are searched for within a radius set
by the orbital curvature (which dominates over LSST's expected astrometric errors). With appropriate use of
kd-trees and similar algorithms for fast searches, only a small fraction (of the order a percent)
of tracklets from the second night need to be examined in detail. The cutoff radius varies
from $\sim$1 arcmin for the case of two consecutive nights to $\sim$1 deg. for a 15-day separation
(as discussed in detail further below). In addition, the velocity of second tracklet is required
to be consistent with the velocity implied by the positions of the two tracklets. After
this step, there are about 10$^{10}$ tracklet pairs for further processing (for $N_w=30$
days).

In the second step, parameters of a parabola (for each coordinate) are constrained using the
positions and velocities of the two tracklets, and this parabolic motion is extended to a third
night to search for matching third tracklet. This step results in up to 10$^{11}$ candidate
tracks.

Using the positions of the three tracklets, parabolic motion (for each coordinate) is fit
in the third step. The velocities implied by this motion are compared to the velocities for
the first and third tracklet. This filtering step reduces the number of candidate tracks
by a factor of about 10$^{5}$ and brings the number of false candidate tracks to
the same range as the number of true tracks close to the Ecliptic. These three matching
and pre-filtering steps bring the number of candidate tracks to a level that
can be easily handled by the IOD filtering step.

We now proceed with a more detailed description of three pre-filtering steps
for candidate tracks.

\subsubsection{The Number of 3-night Combinations in the Search Window}

We can form a candidate triplet of tracklets by first choosing the middle (second) tracklet.
For simplicity, we will measure time of observation in integer days. Given $N_w$ nights
in the search window, the middle tracklet comes from night indexed $k$, with
$2 \le k \le (N_w-1)$. The night that contributes the first tracklet is indexed by $j$,
with $1 \le j \le (k-1)$, and the night that contributes the third tracklet is indexed by $l$,
with $(k+1) \le l \le N_w$. The number of 3-night combinations can be expressed in a closed
form
\begin{equation}
\label{eq:N3}
  N_{3nights} = \sum_{k=2}^{N_w-1} \, (k-1)\, (N_w-k) =\frac{1}{6}N_w^3 - \frac{1}{2}N_w^2 + \frac{1}{3}N_w,
\end{equation}
giving $N_{3nights} = 455$ for $N_w=15$ and $N_{3nights} = 4,060$ for $N_w=30$.  Note
that for large $N_w$, $N_{3night}$ is proportional to $N_w^3$ -- the number of 3-nights
combinations increases by about an order of magnitude when $N_w$ is doubled from
15 days to 30 days.

It is important to point out that in steady-state processing a single night is added to the
window from the previous night, and the first night is dropped. Therefore, only the {\it new}
3-night combinations, where the third night is the last night in the search window, need
be considered in steady-state processing (and the ramp up is easy because of the gradually
increasing search window size). It is straightforward to show that the number of such
3-night combinations is
\begin{equation}
\label{eq:N3n}
  N_{3nights}^{new} = \sum_{k=2}^{N_w-1} \, (k-1) =\frac{1}{2}N_w^2 - \frac{3}{2}N_w + 1,
\end{equation}
yielding $N_{3nights}^{new} = 91$ for $N_w=15$ and $N_{3nights}^{new} = 406$ for $N_w=30$.
Note that $N_{3nights}^{new} \sim N_{3nights} / 10)$ for $N_w=30$, which represents
a significant reduction.

\subsubsection{The Tracklet Motion Vector Accuracy \label{sec:astromerrors}}

In addition to its mean position at the mean epoch, each tracklet constrains the motion vector.
Typical astrometric errors for LSST detections will range from about 50 mas at SNR=100 (dominated
by systematics) to 150 mas at SNR=5 (dominated by random errors). For simplicity, we will assume 
hereafter that the astrometric errors are
$\sigma_a=150$ mas for all detections, or $\sim 100$ mas per coordinate. With a temporal
separation of two detections in a tracklet of $\Delta t$, the motion vector is measured with an
accuracy per coordinate of
\begin{equation}
\label{eq:sigv}
          \sigma_v = 3.6 \, \left({\rm hour} \over \Delta t\right) \,\,\, {\rm arcsec} \, {\rm day}^{-1}.
\end{equation}
With a typical $\Delta t = 30$ min, and assuming a linear motion in each ecliptic coordinate (longitude
$\lambda$ and latitude $\beta$), each coordinate can be predicted at time $t$ with an accuracy of
\begin{equation}
            \sigma_x = 7.2 \, \Delta k \,\,\, {\rm arcsec},
\end{equation}
where $\Delta k$, in days, is the elapsed time between the mean tracklet epoch and time $t$
(for example, the number of nights between the first and the second tracklet in a candidate track).
For illustration, when $\Delta k = 7$ days, $\sigma_x = 50$ arcsec, which is roughly the same
as the typical detection separation in a tracklet, and comparable to typical distance between
two tracklets.  However, it turns out that the positional discrepancies due to a simple linear extrapolation of
motion for NEOs are an order of magnitude larger than the astrometric measurement errors
even in case of two consecutive nights ($\sim$1 arcmin vs. 7 arcsec, respectively). We proceed with
a quantitative analysis of the required matching radius using simulated orbits for main-belt asteroids
and NEOs.

\subsubsection{Initial Linking of Tracklets into Candidate Tracks}

Given a combination of 3 different nights from the search window, for each tracklet
from the first night we can linearly extrapolate its motion vector and require that the
measured  position of a tracklet from the second night is consistent with the predicted
position (the night ordering can be reversed from 1-2-3 to 3-2-1). Given a tracklet from
the first night, it is not necessary to search through all tracklets from the second night.
Search methods such as kd-trees can be used to rapidly reject tracklets that have no
chance of being matched. As an example of a ``poor man's'' rapid search, consider the
fact that tracklets from each night are already ``self-organized'' into about 500 visits,
which correspond to a field of view with a diameter of 3.5 deg. It is easy to show that with
an upper limit on possible motion of 5 deg, only 19 visits from the second night need to be
searched for matching tracklets. This significant reduction of a factor of $\sim$25 in
the number of candidate matching tracklets can be further boosted by applying more
sophisticated tree algorithms.

\begin{figure}[th!]
\centering
\vskip -2.6in
\includegraphics[width=0.95\textwidth]{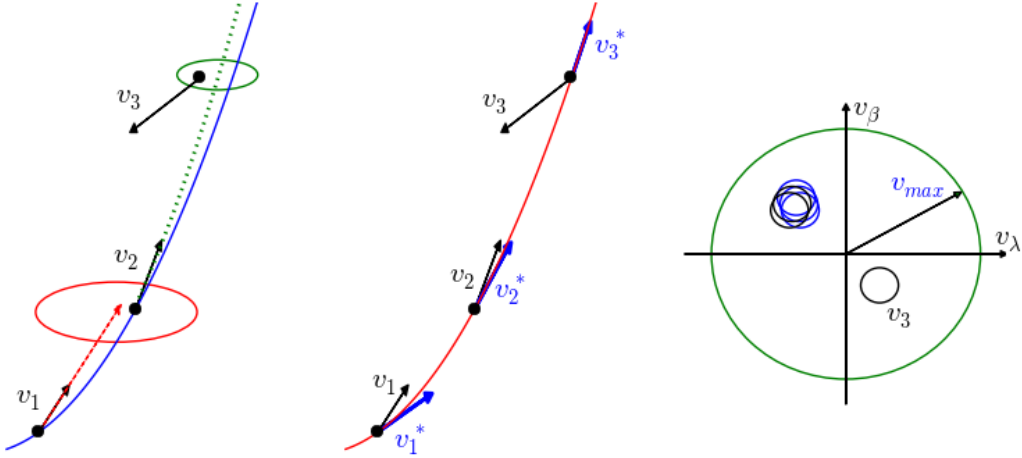}
\vskip -2.7in
\caption{The left panel shows a hypothetical asteroid trajectory as the curved blue line (with
the curvature greatly exaggerated). Three tracklets are shown by the black dots; the
two individual detections per tracklet are not shown, but are implied by the three measured
motion vectors ($v_1$, $v_2$ and $v_3$). The third tracklet illustrates a false tracklet.
The motion vector of the first tracklet is linearly extrapolated to the time of the second
tracklet and matched within the red ellipse. The first two tracklets are then used to
constrain parabolic extrapolation, shown by the green dotted line, which is then matched
within the green ellipse. Given three candidate tracklets, a parabola is fit to their positions
and predicted motion vectors are computed for each tracklet (the blue vectors in the middle
panel). This comparison is illustrated in the right panel, where the circle signifies the cutoff
velocity for forming tracklets. Note that the third tracklet has a measured velocity ($v_3$)
that is inconsistent with the predicted velocity ($v_3^\ast$). The consistency radii are discussed
in the text.
\label{fig:TrackSlide1}}
\end{figure}

Using ecliptic longitude $\lambda$ for illustrative purposes, the predicted search position for the
second tracklet is
\begin{equation}
\label{eq:lambdaPred}
               \lambda_2^\ast = \lambda_1 + v_1^\lambda \, \Delta T_{21},
\end{equation}
where $\Delta T_{21}$ is the elapsed time between the epochs of the first and second tracklet,
and $v_1^\lambda$ is the longitudinal component of $v_1$, the motion vector for the first
tracklet, {\it divided by} cos($\beta$). The expectation value for the number of matches in
an ellipse (see the left panel in Figure~\ref{fig:TrackSlide1})
centered on predicted position ($\lambda_2^\ast, \beta_2^\ast$), and within limits $r_\lambda^{max}$
and  $r_\beta^{max}$ along the Ecliptic longitude and latitude, is given by
\begin{equation}
\label{eq:Nmdk}
     N_{match}(\Delta k) = \pi \, r_\lambda^{max} \, r_\beta^{max}  \, \rho_{tracklet} \left({1 \, {\rm day} \over T_{revisit}}\right)
\end{equation}
where the division by $T_{revisit}$ reflects the fact that each field is revisited on average only
every $T_{revisit}$ days (statistically speaking; the number of matches is zero for all but one
night out of $T_{revisit}$ nights).

The extrapolation given by eq.~\ref{eq:lambdaPred} implies that orbits can be approximated by
linear motion (in each coordinate) over time $\Delta T$. This is an incorrect assumption
due to orbital curvature and we analyze this effect using orbital simulations of MBA and NEO
samples described in \S\ref{sec:MAFdetails}.

Analysis of simulated samples shows that an adequate acceleration limit\footnote{See also
Figure 16 in \citet{LDM-156}.} is $a^{max}=0.02$ deg day$^{-2}$: essentially
all main-belt asteroids and more than 95\% of NEOs satisfy this criterion. If this acceleration
were constant during an interval of $\Delta k$ days, the maximum positional discrepancy
would be proportional to $\Delta k^2$. Numerical analysis of the simulated orbital motions
suggests that an approximately constant selection completeness (as a function of $\Delta k$)
is attained for
\begin{equation}
\label{eq:matching1}
                r_\beta^{max} = A \, \Delta k^{1.5}.
\end{equation}
with $A=1.0$ arcmin, and $r_\lambda^{max} = 5 \, r_\beta^{max}$. The achieved completeness for
a fiducial $\Delta k$=7 days is 0.99 for MBAs and 0.95 for NEOs, with very little dependence
on $\Delta k$ for 1 day $\le \Delta k \le$ 21 days (per single search window -- note that most
objects will have multiple discovery chances).  With this linear motion model, the number
of matched tracklets per single trial tracklet is
\begin{equation}
\label{eq:Nmdk2}
   N^{L}_{match}(\Delta k) = 1.96 \, \left({1 \, {\rm day} \over T_{revisit}}\right) \,
                    \left( \rho_{tracklet}  \over 450 \, {\rm deg}^{-2} \right) \, (\Delta k)^3.
\end{equation}
For example,  the expected number of matches for $\Delta k$=7 days is $\sim$224
(a 19 arcmin by 93 arcmin matching ellipse), and rises to $\sim$6,000 for
$\Delta k$=21 days.

Given the two matched tracklets, we can then approximate the motion as a parabola
\begin{equation}
\label{eq:parabola}
          \lambda(t) = \frac{1}{2}a^\lambda \, t^2 + v^\lambda \, t + \lambda_1,
\end{equation}
where $t = mjd - mjd_1$ (and analogously for latitude $\beta$).  Using the tracklet
positions and the motion vector of the first tracklet, the acceleration can be directly
estimated as
\begin{equation}
 \label{eq:accPred}
             a^\lambda = 2 \, {\lambda_2- \lambda_1 - v_1^\lambda \Delta T_{21} \over \Delta T^2_{21}},
\end{equation}
and the predicted velocity for the second tracklet can be estimated from
\begin{equation}
\label{eq:v2cut}
        (v^\lambda_2)^\ast =  a^\lambda \,\Delta T_{21}  + v_1^\lambda =
      {2(\lambda_2- \lambda_1) \over \Delta T_{21}}  - v_1^\lambda.
\end{equation}

We find that a comparison of $v_2^\lambda$ and $(v^\lambda_2)^\ast$ can further decrease
the number of false tracks (recall \S\ref{sec:falsev}); with tolerances of $\Delta v^\lambda < 0.3$ deg/day and
$\Delta v^\beta < 0.07$ deg/day (applied simultaneously for both coordinates as an
elliptical condition), the reduction is about a factor of 50 (for $v_{max}$ = 1 deg day$^{-1}$),
with only a minimal impact on the sample completeness. Therefore, depending on $\Delta k$, the number of tracklet
pairs per trial tracklet  to continue processing ranges from $\sim$4 for $\Delta k$=7 days to
$\sim$120 for $\Delta k$=21 days. When added over all possible pairs of nights (with
$T_{revisit}=3$ days), the total number of candidate tracklet pairs normalized by
the number of tracklets per night ranges from 350 for $N_w=15$ days to 13,400 for $N_w=30$ days.
Therefore, the following, more involved, selection steps need to be executed for
no more than about $10^{10}$ tracklet pairs (for $N_w=30$ days; and only for $3\times10^{8}$
pairs when $N_w=15$ days). These numbers are significantly lower than the naive estimate of
10$^{15}$ ($10^3\times10^6\times10^6$).

We note that in steady-state processing, the new candidate tracklet pairs need to
be evaluated only for pairs of nights where the second night is the penultimate
night in the search window (all other combinations will have been already computed
on previous days). Because the caching of results from previous night is not
yet implemented in MOPS, we don't account for this reduction (of about a factor of
3 to 6) in the analysis presented here.

Given the acceleration estimate from eq.~\ref{eq:accPred}, the position of the third
tracklet can be predicted from
\begin{equation}
\label{eq:lambdaPred3}
  \lambda_3^\ast = \frac{1}{2} a^\lambda \, \Delta T_{32}^2 + v_2^\lambda \, \Delta T_{32} + \lambda_2.
\end{equation}
Similarly to eq~\ref{eq:matching1}, an approximately constant selection completeness
can be achieved using
\begin{equation}
\label{eq:matching2}
                r_\beta^{max} = B \, \Delta k^{1.5}.
\end{equation}
with $B=0.2$ arcmin, and $r_\lambda^{max} = 5 \, r_\beta^{max}$. Note that the search
area is now 25 times smaller than in the first case, thanks to parabolic rather
than linear extrapolation. Therefore, the number of matched tracklets per single
trial tracklet pair is
\begin{equation}
\label{eq:Nmdk3}
     N^P_{match}(\Delta k) = 0.078 \, \left({1 \, {\rm day} \over T_{revisit}}\right) \,
                    \left( \rho_{tracklet}  \over 450 \, {\rm deg}^{-2} \right) \, (\Delta k)^3,
\end{equation}
and the expected number of matches ranges from 9 for $\Delta k$=7 days to
$\sim$240 for $\Delta k$=21 days.

The total number of candidate tracks per single trial tracklet, for all possible 3-night
combinations (where the third night is the last night in the search window) is
\begin{equation}
\label{eq:Ntt}
   N_{tracklet}^{tracks} = 3.2\times10^{-3} \, \left( \rho_{tracklet}  \over 450 \, {\rm deg}^{-2} \right)^2 \, \left({1 \,{\rm day}\over T_{revisit}}\right)^2 \, \sum_{k=2}^{N_w-1} \sum_{j=1}^{k-1} \, (k-j)^3 \, (N_w-k)^3.
\end{equation}
The normalization constant is equal to $1.96\times0.078\times(\Delta v^\lambda
\Delta v^\beta/v_{max}^2)$, where the term in parenthesis is $\sim0.02$. This
normalization gives the number of candidate tracks per search window normalized
by the number of tracklets per night (which is assumed constant for all nights).
The two terms in the sum reflect the multiplication of the number of matches found
in the first selection step (linear extrapolation from the first to the second night,
eq.~\ref{eq:Nmdk2}) and the number of matches found in the third selection step
(parabolic extrapolation from the second to the third night, eq.~\ref{eq:Nmdk3}).

The sums in eq.~\ref{eq:Ntt} can be evaluated analytically, but the result is cumbersome.
Using numerical evaluation (with $T_{revisit}=3$ days), we find that the number of candidate
tracks per tracklet ranges from $\sim600$ for $N_w=15$ days to $\sim174,000$ for
$N_w=30$ days ($N_{tracklet}^{tracks}$ scales with $N_w^8$ when the third night must be
the last night from the search window).
Therefore, the matching of the candidate third tracklet brings the number of candidate
tracks per search window to the range 10$^{9}$ - 10$^{11}$. The ratio of false candidate tracks
to true tracks is in the range 10$^{3}$ - 10$^{5}$, depending on $N_w$. Despite the reduction by
a factor of  about $10^{10}$ to $10^{12}$ from the combinatorial number of tracklet triplets,
another significant reduction is required before the IOD step can be attempted.

\subsubsection{Using Tracklet Motion Vectors to Prune Candidate Tracks}

The positional matching described above didn't use strong constraints on the tracklet velocities
for the first and third tracklets. Since false tracklets (anything other than a true 
asteroid-asteroid pair of detections) have random velocities, 
velocity filtering can further reduce the number of false tracks.
With three candidate tracklets, a parabolic motion (see eq.~\ref{eq:parabola}) can be fit {\it without}
using tracklet velocities. This fit predicts the velocity of each tracklet from the first derivative of
the fit, which can then be compared to each measured velocity. Figure~\ref{fig:TrackSlide1}
illustrates a situation where, e.g., $v_3$ is inconsistent with the velocity predicted using such
parabolic fit.

The consistency tolerances are driven by the orbital curvature and acceleration, rather than
by the velocity measurement errors (velocities are measured with a precision of about 0.001
deg day$^{-1}$, see eq.~\ref{eq:sigv}). Analysis of the simulated samples described in
\S\ref{sec:MAFdetails} shows that velocity tolerances of $\delta v_\lambda^{max}$=0.12 deg day$^{-1}$
for the longitudinal component and $\delta v_\beta^{max}$=0.03 deg day$^{-1}$ for the latitudinal component
reject most false tracklets with only a few percent effect (per single discovery attempt) on overall
sample completeness.

The probability that a random false tracklet velocity will be consistent with a given expected
velocity is approximately (assuming a uniform distribution of false tracklet velocities)
\begin{equation}
        p_v =  { \delta v_\beta^{max} \, \delta v_\lambda^{max} \over  v_{max}^2 }
\end{equation}
With $v_{max} = 1$ deg day$^{-1}$, $p_v = 0.0036$. In reality, this probability is a bit smaller because
the false tracklet velocity distribution is not uniform (it is biased towards the velocity cutoff).
Finally, the probability that all three tracklets have velocities consistent with those
implied by their positions is $p_v^2 \sim 10^{-5}$ (not $p_v^3$ because $v_2$ was already
subjected to a fairly stringent cut, see eq.~\ref{eq:v2cut}; a more stringent cut here
would provide a reduction by about a factor of five, which we ignore)

This significant reduction in the number of candidate false tracks, due to filtering velocities
of the first and third tracklets, brings it to the range 10$^{4}$ - 10$^{6}$, which is smaller or at
most about the same as the number of true candidate tracks (on the Ecliptic). 

False tracks can also arise from incorrect matches of true (asteroid-asteroid) tracklets 
from different asteroids and these do not have a random distribution
of velocities. We discuss the impact of this below, when evaluating the results 
of numerical (instead of analytical) scaling evaluations, however the increase in
number of false tracks is small, as the density of false detections is much larger than
the density of asteroids.

With this final
reduction, the IOD step can be attempted with no more than about 10$^6$ candidate tracks per
search window.

\subsubsection{The scaling of the number of false tracks with the density of false detections}

The final number of false tracks can be computed using eq.~\ref{eq:Ntt}, after multiplying
the normalization constant by $p_v^2$ to account for velocity filtering. Numerical evaluation
shows that the expected number of false tracks per tracklet can be described as
\begin{equation}
\label{eq:NttFinalFit}
   N_{tracklet}^{tracks} = 2.4 \, \left({N_w\over 30\, {\rm day}}\right)^8 \, \left( \rho_{tracklet}  \over 450
        \, {\rm deg}^{-2} \right)^2 \, \left( {3\,{\rm day} \over T_{revisit}} \right)^2 \,
         \left( {1 \, {\rm deg} \, {\rm day}^{-1}  \over  v_{max} }\right)^6.
\end{equation}
Note the very steep dependence on $N_w$: the large power-law index (8) is a result of the two
powers of 3 under sum in eq.~\ref{eq:Ntt}, and the scaling of the number of three-night
combinations with $N_w^2$ from eq.~\ref{eq:N3n}. The scaling of $N^{falsetracks} = N_{tracklet}^{tracks} \, N_{tracklet}$
with the density of false positive detections is very steep, too. Since $N_{tracklet}$ and $\rho_{tracklet}$
are approximately proportional (in the limit $\rho_{ast}=0$) to $\rho_{FP}^2$ (see eq.~\ref{eq:NttFalse}),
the number of false candidate tracks approximately scales with $\rho_{FP}^6$. 
$N_{tracklet}^{tracks}$ scales with $v_{max}^{-6}$ because of velocity filtering for three tracklets,
with each of the three steps giving a $v_{max}^{-2}$ contribution). 
Since $N^{falsetracks}$ scales with $\rho_{tracklet}^3$, and $\rho_{tracklet}$ scales with $v_{max}^2$
(due to dependence of $A_S$ on $v_{max}$, see eq.~\ref{eq:AS}), $N^{falsetracks}$ is approximately 
independent of $v_{max}$ (but note that the number of intermediate filtering operations does 
depend on $v_{max}$). 

Without the limiting $\rho_{ast}=0$  approximation, eq.~\ref{eq:NttFalse} implies a shallower scaling
of the number of false candidate tracks, $N^{falsetracks}$ with $\rho_{FP}$.
We have determined numerically that the scaling of the number
of false candidate tracks per search window with the density of false detections in difference
images, as well as other relevant parameters, is well described by
\begin{equation}
\label{eq:falsetracks}
   N^{falsetracks} = 4.5 \times 10^6 \, \left( {N_w \over 30 \, {\rm day} } \right)^{8} \left( {\rho_{FP} \over 400 \, {\rm deg}^{-2} }\right)^{3.7}
    \left( {\Delta t  \over 30 \, {\rm min}} \right)^{2.7}
     \left( { 1 \, {\rm deg} \, {\rm day}^{-1} \over v_{max} }  \right)^{1.3}.
\end{equation}
This numerical approximation is only valid around above {\it fiducial values} of various 
parameters\footnote{For example, eq.~\ref{eq:falsetracks} cannot be used to conclude that
an infinitely large $v_{max}$ cutoff would result in no false tracks.}, 
and assumes $\rho_{ast}=100$ deg$^{-2}$ and $T_{revisit}=3$ days. With fiducial parameters, and 
when $\rho_{ast}=0$, the number of false tracklets per night is $\sim10^6$, and the number of false 
tracks per search window with $N_w=30$ days is about 550,000. For $N_w=15$ days,  the number 
of false tracks drops to $\sim2,000$. 

We note that the relevant quantity that determines the number of false candidate tracks is {\it not}
the ratio of false to real (asteroid) detections in difference images, but rather the overall number 
(and density) of false detections.

The scaling result given by eq.~\ref{eq:falsetracks} may prove useful when optimizing cadence
and search strategy, as well as for sizing the required computational resources. For example,
for the rate of 8,200 deg$^{-2}$ false detections from Pan-STARRS1, one would expect a factor
of $7\times10^4$ more false candidate tracks that discussed above (that is, about $10^{11}$).
Even with $N_w$=15 days, the predicted number of false candidate tracks remains of the
order 10$^9$.

Although MOPS algorithms operate in a different way, these analytic probabilistic considerations
explain why the number of candidate tracks produced in MOPS experiments stays approximately
the same (to within a factor of two) even when the number of input tracklets per night is increased
by about an order of magnitude. With $N_{tracklet}^{tracks} \sim2$, the number of  candidate
tracks (both true and false) per search window is about $N^{tracks} \sim 5\times10^6$ for $N_w= 30$ days,
that is, not overwhelmingly larger than the number of true tracks (500,000). In other words, the ratio of
false to true detections of 4:1 generates a ratio of false to true tracklets of 3:1 and a ratio of false to true
candidate tracks of 10:1 (for $N_w=15$ days and $\rho_{ast}=100$ deg$^{-2}$, the ratio of false to true
candidate tracks drops to below 5\%).

This similarity in the number of true and false candidate tracks is in good
agreement with the results of MOPS simulations\footnote{See the top left panel
in Figure 21 in \citet{LDM-156}} (though note that those simulations used more aggressive filtering
based on ``parabolic motion plus topocentric correction'' model, and thus obtained a factor of a few
lower counts of candidate tracks).

\bibliography{neo_capabilities}
\end{document}